\documentclass[12pt]{article}

\usepackage{lscape}
\usepackage{scicite}

\usepackage{amsmath,siunitx}
\usepackage{amssymb}
\usepackage{gensymb}
\usepackage{graphicx}
\usepackage{times}
\usepackage[nottoc]{tocbibind}
\usepackage{titlesec, tabularx, makecell}
\usepackage{float}
\usepackage{sectsty}
\usepackage{caption}
\usepackage{url,color}
\usepackage{mathtools}
\usepackage{rotating}
\usepackage{subcaption}
\usepackage{xc}
\usepackage [english]{babel}
\usepackage [autostyle, english = american]{csquotes}
\usepackage{booktabs,caption}
\usepackage[flushleft]{threeparttable}
\usepackage{graphicx}
\usepackage{adjustbox}
\usepackage{caption}  
\usepackage{rotating}

\usepackage{booktabs} 
\usepackage{threeparttable} 
\usepackage{dcolumn} 
 \newcolumntype{d}[1]{D{.}{.}{#1}}

\usepackage[table]{xcolor}

\definecolor{headinggray}{HTML}{BFBFBF}
\definecolor{rowgray}{HTML}{F2F2F2}
\definecolor{white}{HTML}{FFFFFF}
\definecolor{lightgray}{gray}{0.9}

\newcommand{\sym}[1]{\rlap{#1}}

\usepackage{lscape}
\usepackage{xr}

\usepackage{scicite}

\usepackage{amsmath,siunitx}
\usepackage{amssymb}
\usepackage{gensymb}
\usepackage{graphicx}
\usepackage{times}
\usepackage[nottoc]{tocbibind}
\usepackage{titlesec, tabularx, makecell}
\usepackage{float}
\usepackage{sectsty}
\usepackage{caption}
\usepackage{url,color}
\usepackage{mathtools}
\usepackage{rotating}
\usepackage{subcaption}
\usepackage{xc}
\usepackage [english]{babel}
\usepackage [autostyle, english = american]{csquotes}
\usepackage{booktabs,caption}
\usepackage[flushleft]{threeparttable}
\usepackage{graphicx}
\usepackage{adjustbox}
\usepackage{caption}  
\usepackage{rotating}

\usepackage{booktabs} 
\usepackage{threeparttable} 
\usepackage{scicite}

\usepackage{times}

\newenvironment{sciabstract}{
\begin{quote} \bf}
{\end{quote}}

\title{High-Impact Innovations and Hidden Gender Disparities in Inventor-Evaluator Networks}

\author
{Tara Sowrirajan,$^{1,2}$ Ryan Whalen,$^{3}$ Brian Uzzi$^{1,2\ast}$\\
\\
\normalsize{$^{1}$Kellogg Graduate School of Management, Northwestern University, USA}\\
\normalsize{$^{2}$Northwestern Institute on Complex Systems, Northwestern University, USA}\\
\normalsize{$^{3}$University of Hong Kong Faculty of Law, Hong Kong SAR}\\
\\
\normalsize{$^\ast$Corresponding author:  uzzi@northwestern.edu}
}

\date{}

\begin{document}

\maketitle

\begin{sciabstract}
\normalfont
 We study of millions of scientific, technological, and artistic innovations and find that the innovation gap faced by women is far from universal. No gap exists for conventional innovations. Rather, the gap is pervasively rooted in innovations that combine ideas in unexpected ways – innovations most critical to scientific breakthroughs. Further, at the USPTO we find that female examiners \textit{reject} up to 33 percent more unconventional innovations by women inventors than do male examiners, suggesting that gender discrimination weakly explains this innovation gap. Instead, new data indicate that a configuration of institutional practices explains the innovation gap. These practices compromise the expertise women examiners need to accurately assess unconventional innovations and then “over-assign” women examiners to women innovators, undermining women’s innovations. These institutional impediments negatively impact innovation rates in science but have the virtue of being more amenable to actionable policy changes than does culturally ingrained gender discrimination. 
\end{sciabstract}

\section*{Introduction}

Breakthrough innovations often require unexpected thinking \cite{park_papers_2023, hofstra_diversityinnovation_2020, uzzi_atypical_2013, carpenter_citation_1981, chu_slowed_2021, shi_surprising_2023, fortunato_science_2018, youn_invention_2015, whalen_patent_2020, leahey_what_2023}, an observation summarized in Einstein’s declaration, “we cannot solve our problems with the same thinking we used when we created them.” Yet, despite the breakthrough potential of innovations that employ new thinking, their nature presents a contradiction in capturing their value. By turning away from past ideas, innovations that push against the boundaries of convention seem dubious, raising rather than reducing their rejection rate \cite{uzzi_atypical_2013, leahey_what_2023, foster2015tradition, whalen_boundary_2018}.  Consider the laser - the U.S. Patent office repeatedly rejected the laser’s application for more than three years because its rare combination of ideas from optics and electromagnetism impeded patent examiners ability to recognize its value \cite{bromberg_laser_1991, glover_explaining_2024}. Notwithstanding the laser’s eventual patenting and foundational contributions to other breakthroughs, science policy analysts are raising concerns that the drop in science’s innovativeness may reflect a host of “lost breakthrough innovations” that could address chronic problems and other grand challenges \cite{park_papers_2023, hofstra_diversityinnovation_2020, uzzi_atypical_2013, chu_slowed_2021, shi_surprising_2023, fortunato_science_2018, leahey_what_2023, perkmann_academic_2021, noauthor_darpa-like_2024}.

One policy approach to boost breakthrough rates is to increase the number of women innovators involved in innovation \cite{hofstra_diversityinnovation_2020, page_difference:_2008, woolley_evidence_2010, yang_gender-diverse_2022, chien_improving_2023}. Women are a major and growing proportion of scientifically trained talent, making up about 60\% of college graduates, but their proportional representation among innovators is low, for example making up just 21\% of patented inventors \cite{woolley_evidence_2010, chien_improving_2023, ding_gender_2006, koning_who_2021, nielsen_opinion_2017, toole_progress_2020}. Women would also add new and diverse perspectives into the creative process \cite{shi_surprising_2023, yang_gender-diverse_2022}, as well as the review process as referees, editors, and examiners, which could be more inclusive to unconventional innovations that combine ideas in new or atypical ways \cite{page_difference:_2008}. 

Nevertheless, scarce data inhibits the study and policy analysis of women-created innovations that push the boundaries of conventional thought \cite{gorman_gender_2005}. As such, the espoused benefits of increasing women’s participation in science could be misguided. On the one hand, gender stereotypes could conceivably over-penalize unconventional thinking by new groups of women innovators \cite{chien_improving_2023, ding_gender_2006, lerchenmueller_gender_2018, gorman_gender_2005}, inadvertently lowering innovation rates. On the other hand, more women reviewers could mitigate gender stereotypes, thereby increasing innovation rates \cite{rivera_glass_2021, lee2013bias, zhang_gender_2022}. However, few studies simultaneously examine the production and evaluation sides of the innovation process \cite{gorman_gender_2005}, and no studies do so in connection with innovations that break with conventional thinking. 

Here, we use data on 6.6 million patent applications, as well as information on Patent Office policies obtained via the Freedom of Information Act (FOIA) to study for the first-time links between the gender of innovators and evaluators with respect to high-impact, unconventional innovations. These data allow an in-depth analysis of the size and extent of the gender gap for conventional and unconventional innovations, and new research designs for examining how the gender of who invents and who reviews innovations underlie the gender innovation gap. 

We find that segregating innovations into those building on conventional as opposed to unconventional thinking reveals that the well-documented gender innovation gap is nearly nonexistent for conventional innovations and largest for unconventional innovations. Further, these new data and analyses point to institutional discrimination, not gender discrimination, as a key and actionable explanation for removing barriers to innovation.

\section*{Data}

Our empirical materials include diverse data sources. The three datasets concern scientific and technology patent applications. Patent dataset one includes 6,185,556 patent applications from the U.S. (USPTO) filed from 2001 to 2018. In addition to information about inventors, inventions, and patent application success, we have information about the examiners’ identities and experiences. Patent datasets two and three use Canadian and U.K. patents that contain comparable data to the UPSTO data but fewer applications, 280,128 and 224,365, respectively (see SI S2 for a list of variables).

Across these data we quantify the likelihood of being awarded or denied a patent as a function of the invention’s innovativeness and the innovator’s gender. Following prior work, we define an invention’s level of “unconventionality” depending on how much it incorporates new thinking \cite{park_papers_2023, hofstra_diversityinnovation_2020, uzzi_atypical_2013, carpenter_citation_1981, yang_gender-diverse_2022, kim_technological_2016, arts_paradise_2018}. At one end of the continuum, conventional innovations combine past ideas in familiar ways that have been done before. Unconventional innovations combine past ideas in unexpected ways that have never or only rarely been observed before, thereby pushing the boundaries of accepted thinking \cite{park_papers_2023, uzzi_atypical_2013, youn_invention_2015, leahey2017prominent}. 

The degree to which an innovation incorporates new thinking can be quantified using patent CPC codes \cite{youn_invention_2015, whalen_patent_2020}, which categorically group related ideas into separate technology domains. Methodologically, our unconventionality measure aggregates all pairs of CPC codes over all previous patent applications for each year to compute an observed frequency of CPC code pairings. The measure is annually updated and cumulative to capture changes in activity and relationships among CPC codes, allowing us to compare the unconventionality of inventions from different years. To determine whether the observed frequency of a patent’s CPC code combinations represents conventional or unconventional thinking, we compare a patent application’s observed frequency of each CPC code pair to the frequency expected by random pairing, which gives us a z-score statistic for each pair \cite{youn_invention_2015}. Pairs of CPC codes that occur together less often than expected by chance reflect unexpected combinations of prior ideas. By contrast, pairs of CPC codes that occur together more often than expected by chance reflect conventional combinations of ideas. To get an overall score for an application, we aggregate the z-score pairs (SI S3-4 provide computational details and robustness checks using different methods of aggregation of z-score pairs).

Our measure has been shown to have face validity \cite{youn_invention_2015, whalen_boundary_2018, yang_gender-diverse_2022, pedraza-farina_network_2020}, and in our data we confirmed that it correlates appropriately with other measures of unconventionality \cite{park_papers_2023, whalen_patent_2020} and impact \cite{kim_technological_2016, kaplan_double-edged_2015}. For example, an unconventional patent is 52.1\% more likely than a conventional patent to be in the top one percentile of cited patents ($p < 0.001$) controlling for grant year, team size, and CPC code. Unconventional patents are economically more impactful too as indicated by their inventors being more likely to pay patent maintenance fees ($p < 0.001$) (see SI S12 for computational details). Women’s applications are just as unconventional as men’s in a majority of CPC codes — 69.1\% of CPC codes (Binomial test, $p < 0.0001$) and have statistically similar overall distributions (KS test, $p < 0.05$ on each CPC code). 

\begin{figure}[!ht]
\captionsetup{font=footnotesize}
 \centering
 \includegraphics[width=1\linewidth]{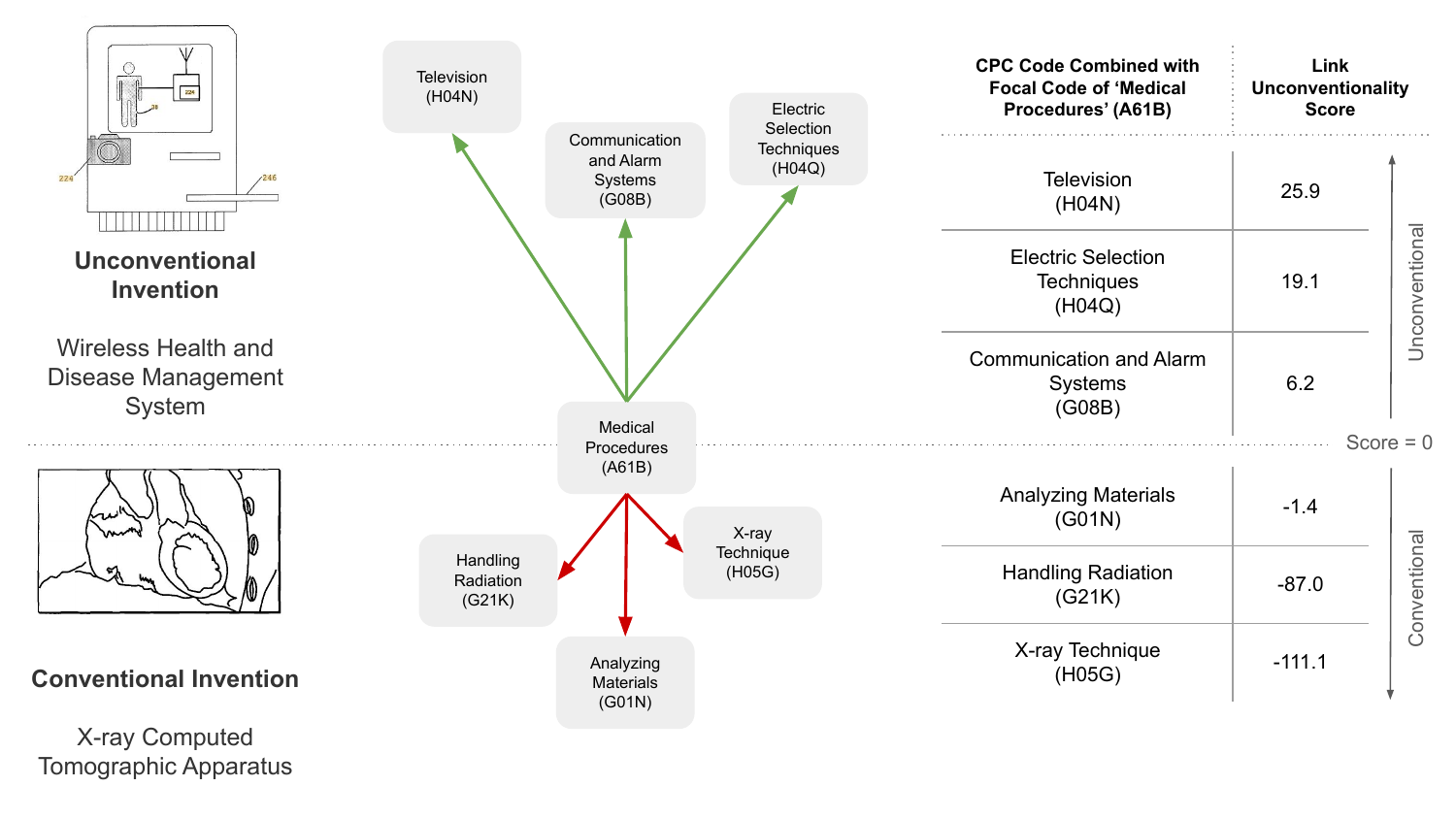}
 \caption{\textbf{Illustrative example of how the same type of innovation can be approached from the perspective of conventional and unconventional thinking.} This example illustrates a comparison between unconventional (top, Pat. App. No. 11/156177) and conventional (bottom, Pat. App. No. 11/367461) inventions, both having the same focal CPC code. It highlights the combined CPC codes and summarizes the invention titles and code description. Network diagrams in the center display combinations of categories in each invention, with proximity indicating commonality and distance indicating rarity. The top network depicts the highly cited unconventional invention “Method and apparatus for health and disease management combining patient data monitoring with wireless internet connectivity,” linking rare pairwise combinations of CPC codes H04N, G08B, and H04Q with focal code A61B. The bottom network depicts the conventional invention “X-ray computed tomographic apparatus, image processing apparatus, and image processing method,” using frequent pairwise combinations of CPC codes G21K, G01N, and H05G with focal code A61B. The table quantifies unconventionality, showing relationships ranging from highly unconventional (positive, green) to conventional (negative, red).}
 \label{fig:fig1Example}
\end{figure}

Figure \ref{fig:fig1Example} provides examples of conventional and unconventional innovations for innovations with similar intended purposes and capabilities to illustrate how the same type of innovation can be approached from the perspective of conventionality or unconventionality. Both innovations were granted patent protection in 2007 and relate to innovations from the same technological domain, “Medical Procedures” (CPC code A61B), and combine the focal technology area (A61B) with three other technological areas (depicted in separate boxes). Figure \ref{fig:fig1Example} lists the unconventionality score between each pair of CPC code categories and the focal category. The length of the arrow between categories is proportional to the level of unconventionality. The unconventional invention combines CPC codes A61B (focal) with H04N, G08B, and H04Q, which are all atypical pairwise combinations with the most uncommon link being between the pair A61B and H04N, combining areas related to medical procedures with information technology, with an unconventionality score of 25.9. The conventional innovation combines CPC codes A61B (focal) with G21K, G01N, and H05G, which are all frequently co-occurring CPC codes pairings that are conventional. The unconventional invention is much more highly cited (citation count of 107) compared to the conventional invention (citation count of 2), exemplifying the higher impact of unconventional innovation.

We followed prior research and used Genderize on names to algorithmically estimate a creator’s gender as man, woman, or undetermined (e.g., only the inventor’s initials are recorded) \cite{yang_gender-diverse_2022}. The USPTO has estimated that 20.3\% of inventors are women, and we find our gender estimation in agreement with this, as Genderize estimated that 20.7\% are women \cite{toole_progress_2020} (SI S1-4 contain computational details and examples).

\section*{Results}

\begin{figure}[!ht]
\captionsetup{font=footnotesize}
 \centering
 \includegraphics[width=1\linewidth]{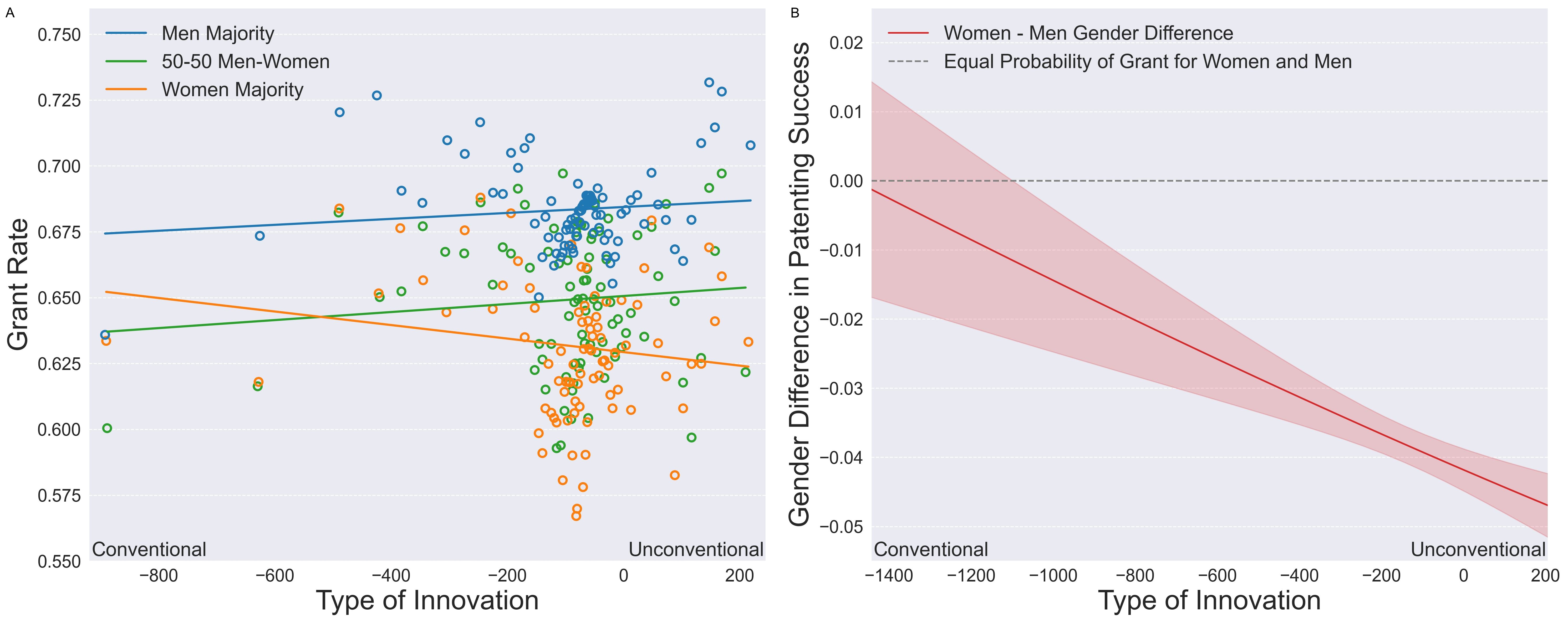}
 \caption{\textbf{The gender gap exists primarily for innovations that incorporate unconventional thinking.} A) This plot shows the relationship between the type of innovation and grant rate by team gender composition in the USPTO data. It depicts a binned scatterplot, showing averages within bins, with regression lines by team gender composition over the full range of data with controls for heterogeneity in CPC codes and team size along with standard errors clustered by year to account for intra-correlation. Women’s success rates decrease as their innovations become increasingly unconventional. B) The residual margins plot (with 95\% CIs) shows the estimated patent grant rate on the interaction of team gender majority and unconventionality from a logit model that includes controls for team size, year, CPC class, examiner gender, examiner and inventor experience, and applicant entity size in the USPTO data. The plot indicates that the innovation gap for women scales with the level of innovation unconventionality. The gender gap appears and widens the more that innovations from women inventors push the boundaries of convention. Innovations that stay within the confines of conventional thinking show less difference in grant rate for women and men innovators. Regression fit and cross validation statistics are reported in the SI and generalizations of the gender gap-unconventionality scaling relationship are reported for international patents.}
 \label{fig:fig2}
\end{figure}

Men and women innovators have starkly divergent patenting experiences regarding conventional and unconventional innovations. Figure \ref{fig:fig2}A depicts the relationship between patenting grant rate and an invention’s unconventionality using the USPTO data, which is binned into 85 quantiles represented by a scatter point of the average grant rate of that bin. The regression line is fit to the scatter points, controlling for team size and 141 separate technological domains as determined by their primary CPC codes with clustered standard errors by year to further control for heterogeneity in the data. The plot shows that when innovations are conventional in nature (indicated by negative z-scores), the difference in the grant rate for men and women innovators is small. Conversely, for unconventional innovations (positive z-scores), the gender innovation gap steadily widens up to over 14.9 percentage points. 

To further understand the link between grant rate, an innovator’s gender, and unconventional innovations, we regressed an application’s probability of being granted on an interaction between the inventor’s gender and the patent’s level of unconventionality along with controls for an inventor’s previous number of patent applications and affiliation (whether part of a corporation), the examiner’s gender and number of previous reviewed patents, and fixed effects for inventor team size, 141 CPC codes, and application year (see SI Table S2 for regression details and SI S5 for robustness to several gender composition operationalizations). Figure \ref{fig:fig2}B is a margins plot summarizing the regression’s estimates. The y-axis measures women’s average probability of patenting success rate minus the men’s rate, with the dashed line showing the point of no gender difference. Values below the line show the women’s innovation gap. 

Contrary to prior work, we find that for highly conventional applications, the well-documented gender gap disappears. Rather, the gender innovation gap appears and enlarges in proportion to the innovation’s unconventionality ($p < 0.001$). In the U.S. alone, we estimate the value of unconventional “lost patents” that would have been granted to women at over \$234 million (1992 USD) by inferring that 2,238 more unconventional patents would be granted if women and men had the same grant rate (see SI S13).
Robustness checks using the U.K. and Canadian datasets indicate that the unconventionality penalty for women innovators is a pervasive phenomenon. U.K. and Canadian women inventors ($p < 0.001$) experience a significant drop in their probability of success when their creations push against conventional thinking (see SI S11 and Table S15 for details). 

\begin{figure}[!htb]
\captionsetup{font=footnotesize}
 \centering
 \includegraphics[width=0.9\linewidth]{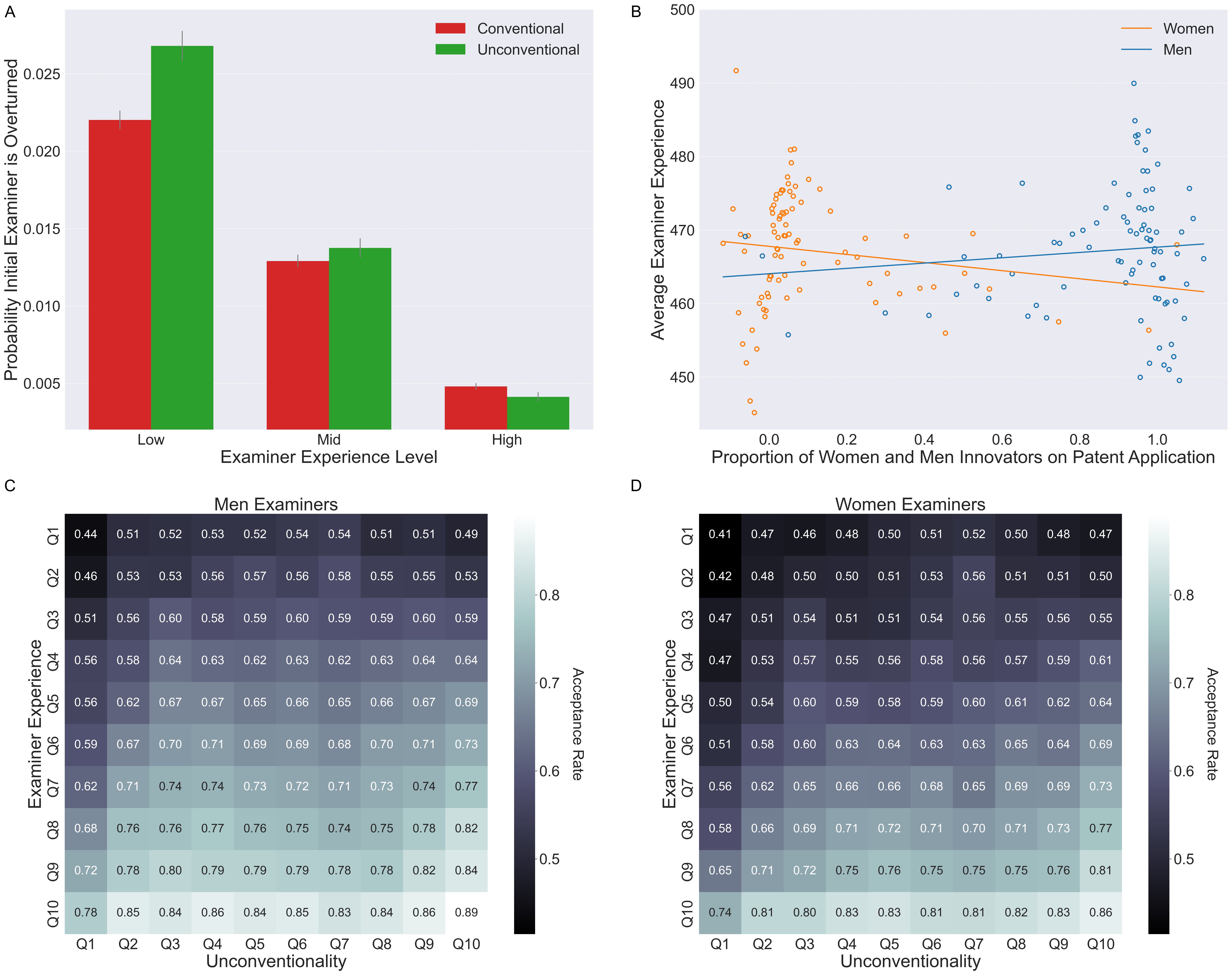}
 \caption{\textbf{Examiner experience and assignment impacts grant rates and decision accuracy, and examiner gender is not a factor in evaluation decisions.} (A) In appeals of patenting decisions, experienced examiners’ decisions are reversed less frequently than inexperienced examiners, suggesting experienced examiners’ decisions are more accurate. For each level of examination experience, the global reversal rate within appealed patent applications for unconventional and conventional innovations is shown. (B) Depicts a binned scatterplot of the proportion of women and men innovators on a patent application and examiner experience (measured in patent application disposals) with regression lines included that control for year, team size, entity, examiner gender, and CPC class. The more women on an application, the less experience their patent examiners have; the more men on an application, the more experience their examiners have. (C) Heatmap of acceptance rate for men examiners by examiner experience quantile and unconventionality quantile (ten quantiles each) on USPTO data. (D) Heatmap of acceptance rate for women examiners by examiner experience quantile and unconventionality quantile (ten quantiles each) on USPTO data. Experience drives openness to unconventionality for both men and women examiners, with similar acceptance rates across experience and unconventionality quantiles.}
 \label{fig:fig3}
\end{figure}

A key explanation for the gender gap has been gender stereotypes \cite{koning_who_2021, gorman_gender_2005, rivera_glass_2021, jensen_gender_2018}. Our evidence, however, is weakly consistent with the gender discrimination claim that men patent examiners are likely to underrate women innovators relative to how women examiners rate women innovators \cite{koning_who_2021}. The USPTO data demonstrates that the opposite is true: women innovators have their highest patenting rates with men examiners and their lowest patenting rates with woman examiners; women examiners are 8.1\% less likely than men examiners to grant patents to women innovators. 

In contrast, our data supports the observation that institutional discrimination, which is rooted in institutional practices, better explains the gender gap. We find that a confluence of institutional factors compromises the expertise women examiners need to accurately assess unconventional innovations, and there is an “over-assignment” of women innovators to women examiners, who are uniformly less likely to grant a patent, especially unconventional patents that require greater expertise to accurately assess.

Examiner expertise is critical for accurate decisions, especially for unconventional innovations that rely on expert judgement to recognize their potential \cite{peer_heuristics_2013}. Our data shows that expertise is positively related to accurate patent decisions. Our first evidence comes from data on appeals of rejected patents, which result in a new panel of expert judges conducting a formal reassessment of the original decision. Figure \ref{fig:fig3}A shows that examiner experience and accurate assessments are positively correlated. In appeals, inexperienced examiners are overturned significantly more than experienced examiners – a relationship that is especially pronounced when they evaluate unconventional innovations. Moreover, Figures \ref{fig:fig3}C and \ref{fig:fig3}D show that across the board, more experienced examiners, irrespective of their gender, have higher grant rates, especially for unconventional inventions ($p < 0.001$, SI Table S9). 

Figure \ref{fig:fig3}B shows that men inventors are assigned significantly more experienced examiners than women inventors, which suggests that women’s unconventional innovations are at higher risk of mistaken rejection because they are assigned to less experienced examiners. Furthermore, Figures \ref{fig:fig3}C and \ref{fig:fig3}D present heatmaps indicating that men and women examiners of equivalent experience accept unconventional patents at the same rate (Pearson correlation coefficient between heatmaps is 0.98, $p < 0.0001$), and that higher examiner experience and higher unconventionality are associated with higher grant rates.

\begin{figure}[!htb]
\captionsetup{font=footnotesize}
 \centering
 \includegraphics[width=0.6\linewidth]{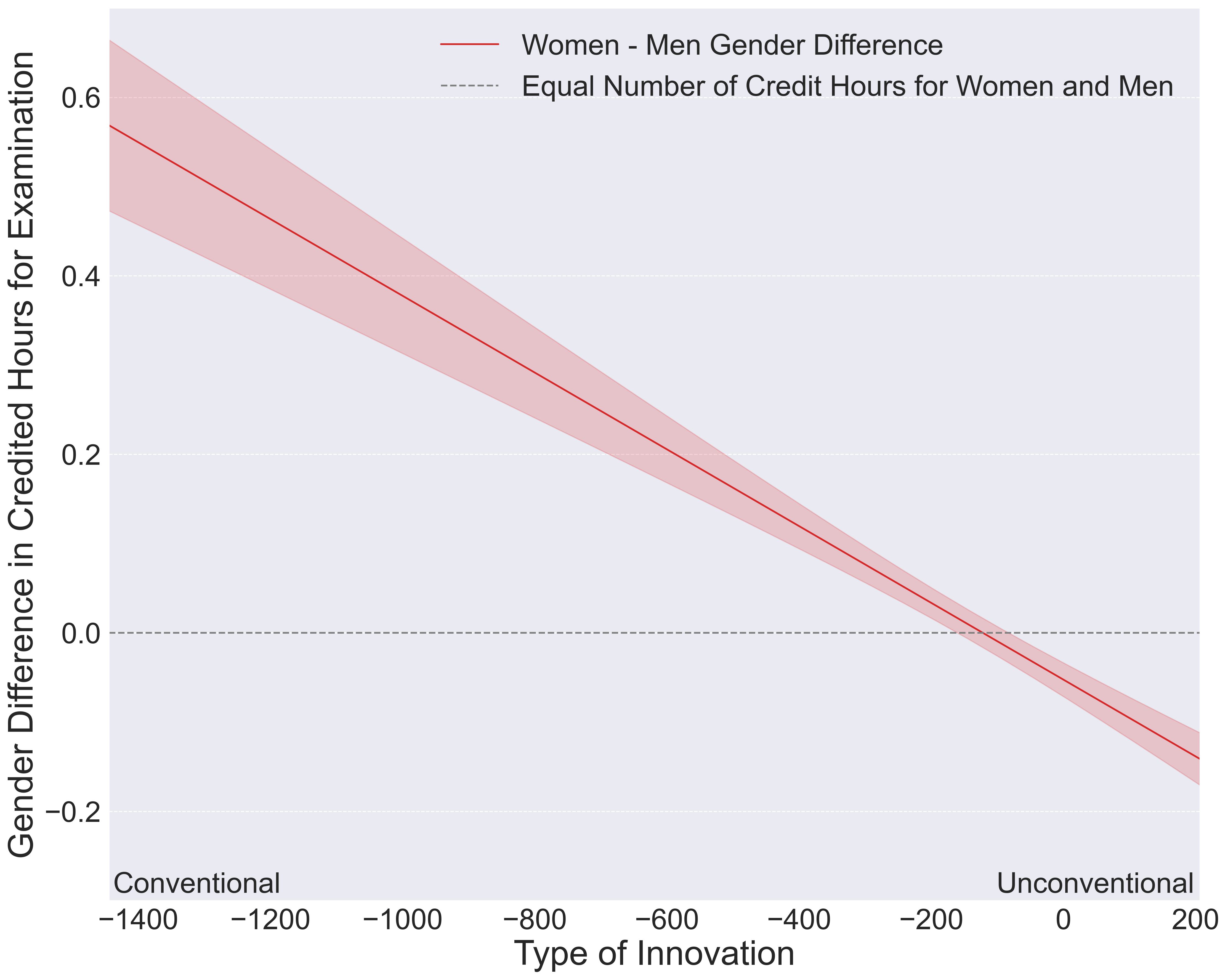}
 \caption{\textbf{Institutional factors disadvantage women inventors with less review time for women’s unconventional innovations.} Analysis of data on patent examination hour crediting by the USPTO shows that the areas women invent in get less time for examination, a relationship that intensifies when women innovators engage in unconventional work. This depicts a residual margins plot (with 95\% CIs) of the estimates of the gender difference in credited examination hours on the interaction of the gender majority of an inventor team and unconventionality from a linear regression model that includes controls for team size, year, CPC class, examiner gender, examiner and inventor experience, and applicant entity size in the USPTO data. }
 \label{fig:fig4}
\end{figure}

The data indicate that institutional arrangements systematically compromise the expertise of women examiners in two ways. First, women examiners on average have less experience than men examiners (T-test $p < 0.0001$) because they drop out of the USPTO workforce earlier than men examiners (SI S10). Second, women examiners receive fewer credit hours than men for their review work, a gap that is especially pronounced when women examine unconventional innovations (SI Table S14). This results in examiners spending less time on applications that are allocated fewer credit hours. New FOIA data reveals that women inventors’ applications typically receive fewer credit hours than male inventors’ applications — a gap that widens for women’s unconventional innovations (Figure \ref{fig:fig4}, see S9 for details). 

Furthermore, other evidence indicates that USPTO practices assign women innovators to the least experienced examiners on average, especially when they engage in unconventional innovation (SI Table S10). The data indicate that women examiners are 16.9\% “over-assigned” to women innovators relative to what is expected by chance (see SI S8 for details). This practice effectively means that on average, women inventors are reviewed more often by less experienced women examiners, who are given less review time than men examiners, accounting for the gender innovation gap for unconventional innovations.

The combined impact of these institutional practices results in fewer breakthrough innovation patent grants from women innovators. Applications from women innovators are over-assigned to women examiners who have lower experience and fewer credit hours to review applications, lowering women inventors’ unconventional innovation grant rates relative to men inventors. Specifically, an all-women inventor team assigned to a woman examiner of low experience has a 37.1\% chance of being granted a patent built on unconventional thinking. Conversely, an all-male inventor team assigned to a male examiner of high experience has an 81.8\% chance of being granted an unconventional patent, representing a stark 44.7 percentage point gender gap in success. Given the relationships in the data, if only half of the unconventional patent applications from women majority teams that were originally assigned to low experienced, women examiners were reassigned to women examiners with high experience, this could conceivably result in 13\% more potentially high impact innovations (SI S13.1).

\section*{Discussion}

Unconventional innovations are especially important to scientific advancement and cultural expression \cite{park_papers_2023, uzzi_atypical_2013, leahey_what_2023}. By pushing against the boundaries of convention, these innovations are more likely to generate scientific and technical breakthroughs, providing new solutions to intractable problems and introducing unforeseen product categories. In science and technology, we found a common denominator that explains the differences in the success of men and women innovators. When innovation respects the boundaries of convention, building on known and taken-for-granted combinations of existing innovations and ideas, men and women innovators have equal chances of the successful adoption of their ideas and inventions. Conversely, when innovation pushes against the boundaries of convention, combining pre-existing innovations and ideas in ways that are never or rarely seen before, men are increasingly rewarded while women are increasingly penalized. This is true even though women and men are equally likely to attempt unconventional innovation.

Contrary to prior work that has explained patenting differences with an emphasis on gender stereotypes \cite{koning_who_2021, ross_women_2022}, the data indicate that women innovators have their highest patenting rates when assigned to a man examiner. Evidence suggests that the difference in successfully patenting unconventional innovations is explained by a confluence of institutional practices that lower women examiners’ expertise and over-assign them to women inventors, not by gender stereotypes. The higher rejection rate for women’s unconventional innovations is explained by the systematic assignment of less experienced examiners who are given relatively less time to review women’s applications. The examiner experience gap means patent applications by men inventors are more likely to be examined by experienced examiners who grant applications more often and are more likely to make accurate decisions. 

Future study could expand beyond the context of science and technology studied here to assess the universality of this phenomenon. 

Our work has theoretical and policy implications for the gender innovation gap. The unconventionality penalty that women innovators experience defines who is allowed to reach for the stars, and it has broad implications for innovation policy. Many proposals aimed at ameliorating the more general gender innovation gap focus on increasing the number of women creators \cite{toole_progress_2020, blickenstaff_women_2005} with reason, but our work focuses points of policy on unconventional innovators who are more likely to address chronic, impactful problems. 

First, by identifying institutional practices that worsen outcomes for women innovators of innovations that break with conventions, our findings help move the innovation gender gap discussion beyond difficult-to-address challenges, such as culturally ingrained stereotyping and bias, to more concrete policy responses that may help close the gap. While gender stereotypes affect gender inequality \cite{gorman_gender_2005, rivera_glass_2021}, intrapsychic biases are difficult to address. They can arise well before women become innovators, by factors outside the control of organizations, and are apt to change slowly. A focus on institutional arrangements complements work on stereotypes with interventions that allow for measurable and potentially immediate changes in outcomes. This draws attention not only to raising innovation rates by understanding the conditions that help inventors innovate successfully, but also to helping evaluators of innovation recognize the merits of innovations that break with conventional ideas \cite{whalen_boundary_2018, peng_promotional_2024}. 

Second, the findings also raise the policy speculation that institutional practices employed to decrease gender discrimination, such as gender matching of innovators and examiners, may ironically result in unintended consequences, negating the practice’s potential benefits. The USPTO and other analogous assessment bodies could address this issue by striving not just for diverse gender representation amongst assessors, but also by trying to ensure that assessors acquire equivalent experience and are distributed across groups of applicants in ways that address multiple assignment biases. 

Third, other institutions that play a vital role as gatekeepers of innovative success should ensure gender parity in resources allocated to making those decisions, such as the hours that examiners are credited for their work. They could also provide more resources for innovations that push against the boundaries of convention. In conjunction, new administrative policies may help ameliorate the gender innovation gap by helping facilitate longer careers for women, both on the side of creators and evaluators. Moreover, systematic efforts to expose junior evaluators to unconventional creations and train them in how to assess unconventional innovations, which have a sparse historical record from which to extrapolate, may help reduce the grant rate gap associated with experience and consequently reduce the gender innovation gap. Relatedly, in conjunction with experience, the degree to which patent examiners receive sufficient time to assess the merits of unconventional applications may affect grant rates. Providing additional review time and training resources could partly make up for a deficit of adequate expertise. Broadly, future research that focuses on gender should strive to better understand how gender correlates with an array of factors; otherwise, a focus on overcoming one issue may only exacerbate other issues.

A limitation of our research is that the gender–experience difference and innovator–assessor gender matching mechanisms we identify here were pervasive and global but necessarily limited to the context where we had the richest data—the Patent Office. In this context, the centralized organization of the system may prove more amenable to reform than other creative domains. While the USPTO Director can institute reforms to address biases in the patent examination system, successfully reforming the less bureaucratic assessment systems may prove more elusive. However, it remains a potentially informative line of future research that could help develop policy aimed at closing the gender innovation gap and reducing barriers to recognizing the merits of innovations that break with convention, both of which can aid in increasing innovation and breakthroughs.

\begingroup
\renewcommand{\addcontentsline}[3]{}
\bibliographystyle{Science}
\bibliography{bib}

\begin{thebibliography}{10}

\bibitem{park_papers_2023}
M.~Park, E.~Leahey, R.~J. Funk, Papers and patents are becoming less disruptive over time, {\it Nature\/} {\bf 613}, 138--144 (2023).

\bibitem{hofstra_diversityinnovation_2020}
B.~Hofstra, {\it et~al.\/}, The {Diversity}–{Innovation} {Paradox} in {Science}, {\it Proceedings of the National Academy of Sciences\/} {\bf 117}, 9284--9291 (2020).

\bibitem{uzzi_atypical_2013}
B.~Uzzi, S.~Mukherjee, M.~Stringer, B.~Jones, Atypical {Combinations} and {Scientific} {Impact}, {\it Science\/} {\bf 342}, 468--472 (2013).

\bibitem{carpenter_citation_1981}
M.~P. Carpenter, F.~Narin, P.~Woolf, Citation rates to technologically important patents, {\it World Patent Information\/} {\bf 3}, 160--163 (1981).

\bibitem{chu_slowed_2021}
J.~S.~G. Chu, J.~A. Evans, Slowed canonical progress in large fields of science, {\it Proceedings of the National Academy of Sciences\/} {\bf 118}, e2021636118 (2021).

\bibitem{shi_surprising_2023}
F.~Shi, J.~Evans, Surprising combinations of research contents and contexts are related to impact and emerge with scientific outsiders from distant disciplines, {\it Nature Communications\/} {\bf 14}, 1641 (2023).

\bibitem{fortunato_science_2018}
S.~Fortunato, {\it et~al.\/}, Science of science, {\it Science\/} {\bf 359} (2018).

\bibitem{youn_invention_2015}
H.~Youn, D.~Strumsky, L.~M.~A. Bettencourt, J.~Lobo, Invention as a combinatorial process: evidence from {US} patents, {\it Journal of The Royal Society Interface\/} {\bf 12} (2015).

\bibitem{whalen_patent_2020}
R.~Whalen, A.~Lungeanu, L.~DeChurch, N.~Contractor, Patent {Similarity} {Data} and {Innovation} {Metrics}, {\it Journal of Empirical Legal Studies\/} {\bf 17}, 615--639 (2020).

\bibitem{leahey_what_2023}
E.~Leahey, J.~Lee, R.~J. Funk, What {Types} of {Novelty} {Are} {Most} {Disruptive}?, {\it American Sociological Review\/} {\bf 88}, 562--597 (2023).

\bibitem{foster2015tradition}
J.~G. Foster, A.~Rzhetsky, J.~A. Evans, Tradition and innovation in scientists’ research strategies, {\it American sociological review\/} {\bf 80}, 875--908 (2015).

\bibitem{whalen_boundary_2018}
R.~Whalen, Boundary {Spanning} {Innovation} and the {Patent} {System}: {Interdisciplinary} {Challenges} for a {Specialized} {Examination} {System}, {\it Research Policy\/} {\bf 47}, 1334--1343 (2018).

\bibitem{bromberg_laser_1991}
J.~L. Bromberg, {\it The laser in {America}, 1950-1970\/} (MIT press, 1991).

\bibitem{glover_explaining_2024}
J.~Glover, Explaining novel scientific concepts to people whose technical acumen does not extend to turning it off, then turning it on again, {\it Nature\/}  (2024).

\bibitem{perkmann_academic_2021}
M.~Perkmann, R.~Salandra, V.~Tartari, M.~McKelvey, A.~Hughes, Academic engagement: {A} review of the literature 2011-2019, {\it Research Policy\/} {\bf 50}, 104114 (2021).

\bibitem{noauthor_darpa-like_2024}
A {DARPA}-like agency could boost {EU} innovation — but cannot come at the expense of existing schemes, {\it Nature\/} {\bf 629}, 504--504 (2024).

\bibitem{page_difference:_2008}
S.~Page, {\it The {Difference}: {How} the {Power} of {Diversity} {Creates} {Better} {Groups}, {Firms}, {Schools}, and {Societies}\/} (Princeton University Press, Princeton, N.J.; Woodstock, 2008), new edition with a new preface by the author edition edn.

\bibitem{woolley_evidence_2010}
A.~W. Woolley, C.~F. Chabris, A.~Pentland, N.~Hashmi, T.~W. Malone, Evidence for a {Collective} {Intelligence} {Factor} in the {Performance} of {Human} {Groups}, {\it Science\/} {\bf 330}, 686--688 (2010).

\bibitem{yang_gender-diverse_2022}
Y.~Yang, T.~Y. Tian, T.~K. Woodruff, B.~F. Jones, B.~Uzzi, Gender-diverse teams produce more novel and higher-impact scientific ideas, {\it Proceedings of the National Academy of Sciences\/} {\bf 119}, e2200841119 (2022).

\bibitem{chien_improving_2023}
C.~V. Chien, L.~L. Ouellette, Improving equity in patent inventorship, {\it Science\/} {\bf 382}, 1128--1129 (2023).

\bibitem{ding_gender_2006}
W.~W. Ding, F.~Murray, T.~E. Stuart, Gender {Differences} in {Patenting} in the {Academic} {Life} {Sciences}, {\it Science\/} {\bf 313}, 665--667 (2006).

\bibitem{koning_who_2021}
R.~Koning, S.~Samila, J.-P. Ferguson, Who do we invent for? {Patents} by women focus more on women’s health, but few women get to invent, {\it Science\/} {\bf 372}, 1345--1348 (2021).

\bibitem{nielsen_opinion_2017}
M.~W. Nielsen, {\it et~al.\/}, Opinion: {Gender} diversity leads to better science (vol 114, pg 1740, 2017), {\it Proceedings of the National Academy of Sciences of the United States of America\/} {\bf 114}, E2796--E2796 (2017).

\bibitem{toole_progress_2020}
A.~A. Toole, {\it et~al.\/}, Progress and {Potential} 2020 update on {U}.{S}. women inventor-patentees, {\it USPTO - Office of the Chief Economist - IP Data Highlights\/} p.~12 (2020).

\bibitem{gorman_gender_2005}
E.~H. Gorman, Gender {Stereotypes}, {Same}-{Gender} {Preferences}, and {Organizational} {Variation} in the {Hiring} of {Women}: {Evidence} from {Law} {Firms}, {\it American Sociological Review\/} {\bf 70}, 702--728 (2005).

\bibitem{lerchenmueller_gender_2018}
M.~J. Lerchenmueller, O.~Sorenson, The gender gap in early career transitions in the life sciences, {\it Research Policy\/} {\bf 47}, 1007--1017 (2018).

\bibitem{rivera_glass_2021}
L.~A. Rivera, J.~Owens, Glass {Floors} and {Glass} {Ceilings}: {Sex} {Homophily} and {Heterophily} in {Job} {Interviews}, {\it Social Forces\/} {\bf 99}, 1363--1393 (2021).

\bibitem{lee2013bias}
C.~J. Lee, C.~R. Sugimoto, G.~Zhang, B.~Cronin, Bias in peer review, {\it Journal of the American Society for information Science and Technology\/} {\bf 64}, 2--17 (2013).

\bibitem{zhang_gender_2022}
L.~Zhang, Y.~Shang, Y.~Huang, G.~Sivertsen, Gender differences among active reviewers: an investigation based on publons, {\it Scientometrics\/} {\bf 127}, 145--179 (2022).

\bibitem{kim_technological_2016}
D.~Kim, D.~B. Cerigo, H.~Jeong, H.~Youn, Technological novelty profile and invention’s future impact, {\it EPJ Data Science\/} {\bf 5}, 1--15 (2016).

\bibitem{arts_paradise_2018}
S.~Arts, L.~Fleming, Paradise of {Novelty}—{Or} {Loss} of {Human} {Capital}? {Exploring} {New} {Fields} and {Inventive} {Output}, {\it Organization Science\/}  (2018).

\bibitem{leahey2017prominent}
E.~Leahey, C.~M. Beckman, T.~L. Stanko, Prominent but less productive: The impact of interdisciplinarity on scientists’ research, {\it Administrative Science Quarterly\/} {\bf 62}, 105--139 (2017).

\bibitem{pedraza-farina_network_2020}
L.~G. Pedraza-Fariña, R.~Whalen, A {Network} {Theory} of {Patentability}, {\it The University of Chicago Law Review\/} {\bf 87} (2020).

\bibitem{kaplan_double-edged_2015}
S.~Kaplan, K.~Vakili, The double-edged sword of recombination in breakthrough innovation, {\it Strategic Management Journal\/} {\bf 36}, 1435--1457 (2015).

\bibitem{jensen_gender_2018}
K.~Jensen, B.~Kovacs, O.~Sorenson, Gender differences in obtaining and maintaining patent rights, {\it Nature Biotechnology\/} {\bf 36}, 307-- (2018).

\bibitem{peer_heuristics_2013}
E.~Peer, E.~Gamliel, Heuristics and {Biases} in {Judicial} {Decisions}, {\it Court Review\/} {\bf 49}, 114--119 (2013).

\bibitem{ross_women_2022}
M.~B. Ross, {\it et~al.\/}, Women are {Credited} {Less} in {Science} than are {Men}, {\it Nature\/} pp. 1--2 (2022).

\bibitem{blickenstaff_women_2005}
J.~C. Blickenstaff, Women and science careers: leaky pipeline or gender filter?, {\it Gender and Education\/} {\bf 17}, 369--386 (2005).

\bibitem{peng_promotional_2024}
H.~Peng, H.~S. Qiu, H.~B. Fosse, B.~Uzzi, Promotional language and the adoption of innovative ideas in science, {\it Proceedings of the National Academy of Sciences\/} {\bf 121}, e2320066121 (2024).

\bibitem{genderize}
genderize.io, \url{https://genderize.io/}. Accessed: 2022-06-30.

\bibitem{ahmadpoor2017dual}
M.~Ahmadpoor, B.~F. Jones, The dual frontier: Patented inventions and prior scientific advance, {\it Science\/} {\bf 357}, 583--587 (2017).

\bibitem{albert1991direct}
M.~B. Albert, D.~Avery, F.~Narin, P.~McAllister, Direct validation of citation counts as indicators of industrially important patents, {\it Research policy\/} {\bf 20}, 251--259 (1991).

\bibitem{bessen2008value}
J.~Bessen, The value of us patents by owner and patent characteristics, {\it Research Policy\/} {\bf 37}, 932--945 (2008).

\bibitem{cimpian2020understanding}
J.~R. Cimpian, T.~H. Kim, Z.~T. McDermott, Understanding persistent gender gaps in stem, {\it Science\/} {\bf 368}, 1317--1319 (2020).

\bibitem{fischer2014testing}
T.~Fischer, J.~Leidinger, Testing patent value indicators on directly observed patent value—an empirical analysis of ocean tomo patent auctions, {\it Research policy\/} {\bf 43}, 519--529 (2014).

\bibitem{frakes2015patent}
M.~D. Frakes, M.~F. Wasserman, Patent office cohorts, {\it Duke LJ\/} {\bf 65}, 1601 (2015).

\bibitem{frakes2017time}
M.~D. Frakes, M.~F. Wasserman, Is the time allocated to review patent applications inducing examiners to grant invalid patents? evidence from microlevel application data, {\it Review of Economics and Statistics\/} {\bf 99}, 550--563 (2017).

\bibitem{hall2005market}
B.~H. Hall, A.~Jaffe, M.~Trajtenberg, Market value and patent citations, {\it RAND Journal of economics\/} pp. 16--38 (2005).

\bibitem{harhoff1999citation}
D.~Harhoff, F.~Narin, F.~M. Scherer, K.~Vopel, Citation frequency and the value of patented inventions, {\it Review of Economics and statistics\/} {\bf 81}, 511--515 (1999).

\bibitem{harhoff2009duration}
D.~Harhoff, S.~Wagner, The duration of patent examination at the european patent office, {\it Management Science\/} {\bf 55}, 1969--1984 (2009).

\bibitem{karimi2016inferring}
F.~Karimi, C.~Wagner, F.~Lemmerich, M.~Jadidi, M.~Strohmaier, {\it Proceedings of the 25th International conference companion on World Wide Web\/} (2016), pp. 53--54.

\bibitem{li2021share}
S.~Li, Y.~Li, J.~Ni, J.~McAuley, Share: a system for hierarchical assistive recipe editing, {\it arXiv preprint arXiv:2105.08185\/}  (2021).

\bibitem{majumder-etal-2019-generating}
B.~P. Majumder, S.~Li, J.~Ni, J.~McAuley, {\it Proceedings of the 2019 Conference on Empirical Methods in Natural Language Processing and the 9th International Joint Conference on Natural Language Processing (EMNLP-IJCNLP)\/} (Association for Computational Linguistics, Hong Kong, China, 2019), pp. 5976--5982.

\bibitem{medeiros2016fintech}
M.~Medeiros, B.~Chau, Fintech-stake a patent claim?, {\it Intellectual Property Journal\/} {\bf 28}, 303 (2016).

\bibitem{moser2015patent}
P.~Moser, J.~Ohmstedt, P.~W. Rhode, Patent citations and the size of the inventive step-evidence from hybrid corn, {\it Tech. rep.\/}, National Bureau of Economic Research (2015).

\bibitem{santamaria2018comparison}
L.~Santamar{\'\i}a, H.~Mihaljevi{\'c}, Comparison and benchmark of name-to-gender inference services, {\it PeerJ Computer Science\/} {\bf 4}, e156 (2018).

\bibitem{toole2019progress}
A.~A. Toole, {\it et~al.\/}, Progress and potential, {\it IP Data Highlights\/} {\bf 20} (2019).

\bibitem{trajtenberg1990penny}
M.~Trajtenberg, A penny for your quotes: patent citations and the value of innovations, {\it The Rand journal of economics\/} pp. 172--187 (1990).

\end{thebibliography}
\nocite{*}
\endgroup

\cleardoublepage

\begin{center}
    \LARGE Supplementary Information for\\[2ex]\textbf{High-Impact Innovations and Hidden Gender Disparities in Inventor-Evaluator Networks}
\end{center}

\begin{center}
    \large
    Tara Sowrirajan,$^{1,2}$ Ryan Whalen,$^{3}$ Brian Uzzi$^{1,2\ast}$\\
    \normalsize
    \vspace{2ex}
    {$^{1}$Kellogg Graduate School of Management, Northwestern University, USA}\\
    {$^{2}$Northwestern Institute on Complex Systems, Northwestern University, USA}\\
    {$^{3}$University of Hong Kong Faculty of Law, Hong Kong SAR}\\
    \vspace{2ex}
    {$^\ast$Corresponding author: uzzi@northwestern.edu}
\end{center}

\vspace{4cm}
{\raggedright
{\bf This PDF file includes:}\\

Data and Methods\\

Tables S1 to S18\\

Findings \\

Figures S1 to S3 \\
}

\pagebreak

\renewcommand{\thetable}{S\arabic{table}}
\renewcommand{\thesection}{S\arabic{section}}
\renewcommand{\thesubsection}{\thesection.\arabic{subsection}}
\renewcommand{\thefigure}{S\arabic{figure}} 
\renewcommand{\theequation}{\arabic{equation}} 
\makeatletter 
\def\tagform@#1{\maketag@@@{(S\ignorespaces#1\unskip\@@italiccorr)}}
\makeatother

\newcommand{\citenumfont}[1]{\textit{S#1}}
\makeatletter
\renewcommand{\l@section}{\@dottedtocline{1}{1.5em}{2.6em}}
\renewcommand{\l@subsection}{\@dottedtocline{2}{4.0em}{3.6em}}
\renewcommand{\l@subsubsection}{\@dottedtocline{3}{7.4em}{4.5em}}
\makeatother

\pagebreak

\maketitle
\tableofcontents
\pagebreak

\addcontentsline{toc}{section}{Data and Methods}
\part*{Data and Methods}

\section{Data Description}
This work makes use of three sources of data spanning multiple diverse innovation contexts.
This includes international patent data from three different countries: the USA, the UK, and Canada.

\subsection{International Patent Application Data}
Three datasets focus on science and technology and include the available universe of over 6.6 million U.S., U.K, and Canadian patent applications. 
The patent data from the US consists of 6,185,556 patents filed from the years 2001 to 2018. 
The patent data from the UK consists of 224,365 patents filed from the years 1985 to 2018. 
The patent data from Canada consists of 280,128 patents filed from the years 1990 to 2020. 

The patent data from the USPTO is the richest and includes diverse features regarding inventors on each application, the examiner, whether or not a large entity is the assignee, the CPC codes (technical classification) of the invention, the status of the patent application, the set of patent applications that are appealed and their revised decisions, credit hours assigned to each CPC code, and transactions between patent inventors and examiners amongst other features. 
The patent application data from the UK and Canada are similar in that they have features such as CPC codes, application status, inventors, and filing year, but are less feature-rich compared to the USPTO data.

In this study, a network-based measure of unconventionality was developed to quantify the extent of boundary-spanning among technical areas (CPC codes) combined in a patent. To calculate this measure, a patent must have at least two technical areas listed. Moreover, a baseline network of technical areas was constructed using the first year of patent data from 2001. In subsequent analyses, we focused on a subset of patents that had at least two CPC codes from the year 2002 onwards, which comprised a total of 392,791 patent applications from the US.

To be consistent with the US patent data, which comprises our most extensive and valuable collection of data, we focus on a subset of the patent application data from the UK and Canada by limiting the filing year range to the years 2002 to 2018 and the number of CPC codes to having at least two CPC codes listed at the subclass level of the CPC hierarchy. We also focus on this subset based on filing year and number of CPC codes regarding patent applications for the US for analysis involving the degree of unconventionality of an invention. 
For patent applications filed in the years 2002 to 2018 with at least two CPC codes, Canada comprised 82,896 applications and the UK comprised 66,709 applications.

\section{Operationalization of Variables}

We use three different datasets. Across them, we compute metrics for unconventionality, gender composition, and outcome variables relating to acceptance. 

\subsection{Gender}
We use a gender classification algorithm called genderize (from genderize.io) to predict the gender of a person given their first name. Each name is classified a gender with an associated probability that quantifies the certainty of the assigned gender \cite{genderize}. The three classes are male, female, and unknown gender. 

This automated approach has several advantages. It allows us to impute gender to a much larger number of inventors than would be possible manually. It also corresponds to methods used both in the scientific literature \cite{ross_women_2022, yang_gender-diverse_2022}, and by the USPTO itself to describe and analyze the patenting gender gap\cite{toole_progress_2020, toole2019progress}. Our estimated gender ratios largely agree with those published by the USPTO in its studies of the patenting gender gap. For instance, as of 2016 the USPTO estimated that 20.7\% of US-based granted patents had at least one woman inventor \cite{toole2019progress}. By comparison, our method estimates of this share of patents was 20.3\%, within 0.37 percentage points of the reported gender statistics. Name-to-gender inference algorithms are capable of high precision \cite{santamaria2018comparison, karimi2016inferring}, but they are not without limitations—both because of ambiguity in relation to some names and because gender is a non-binary and subjective state. To address the possible impact of misclassification, we perform analyses with inventor gender measured in three distinct ways—as inferred either women or men majority teams, with inferred single-gender majority teams and 50\% Men-50\% Women teams, and all women or men teams. 

\subsection{Different Metrics for Gender Composition}
To be consistent with the literature in this area \cite{koning_who_2021, yang_gender-diverse_2022}, we utilize the measures regarding majority female ($n_{female}/n_{total} \geq 0.5$) and majority male ($n_{male}/n_{total} > 0.5$), mixed teams ($n_{female}/n_{total} = n_{male}/n_{total} = 0.5$), as well as single-gender teams ($n_{gender}/n_{total} = 1$), where $n$ refers to the count of a particular gender. We also utilize the gender of the first listed inventor in several analyses. 

As a robustness check, we also measure inventor team gender composition in two additional ways. We use a gender operationalization indicating whether a team is majority female, majority male, or 50\% female and 50\% male, as well as a gender operationalization indicating whether the inventor team is all female or not. These variables are used in regressions as depicted in Tables \ref{regFirstAuth} and \ref{regMajority}.

\subsection{Patent Data}

We utilize several salient features in logistic regression models to predict patent acceptance. 
These are broadly depicted in Table \ref{tab:variablesPatent}. 

\begin{table}[!htbp]
\caption{\textbf{Patent Data Variable Operationalization}}
\label{tab:variablesPatent}
\begin{tabular}{>{\raggedright\arraybackslash}p{0.4\linewidth}|>{\raggedright\arraybackslash}p{0.6\linewidth}}

 \rowcolor{gray!20}
 \textbf{Variable} & \textbf{Operationalization} \\ 
 \hline
 Patent Acceptance & Granted = 1, Abandoned = 0 \\ 
 \rowcolor{gray!10}
 Proportion Female Inventors & $n_{female}/n_{total}$ \\

 Female Majority Inventor Team & 1 if $n_{female}/n_{total} >=0.5$, else 0 if $n_{male}/n_{total} >0.5$\\ 
 \rowcolor{gray!10}
 Team Gender Composition & Men Majority = 0, Women Majority = 1, 50\% Men-50\% Women = 2 \\ 
 
All-Female Inventor Team & 1 if $n_{female}/n_{total} = 1$, else 0 \\ 
 \rowcolor{gray!10}
All-Male Inventor Team & 1 if $n_{male}/n_{total} = 1$, else 0 \\ 
 Unconventionality Level & Z-Score metric: $>0$ unconventional, $<0$ conventional \\ 
 \rowcolor{gray!10}
 Big Entity & 1 if entity is big, else 0 \\ 
 Female Examiner & Female = 1, else 0 \\ 
 \rowcolor{gray!10}
 Average Inventor Experience & Average number of patent applications submitted by the inventors on team \\ 
 Examiner Experience & Number of previous patent applications examined \\ 
 \rowcolor{gray!10}
  Examiner Experience Level & Low, Mid, and High Level of Examiner Experience cut into three quantiles \\ 
 Team Size & Number of inventors, right-capped \\
\rowcolor{gray!10}
 Year & Patent applications from the years 2001-2018 for the USPTO, focus on years 2002-2018 for unconventionality analysis internationally \\ 
 
 CPC Code & Patent technical area at class level \\
  \rowcolor{gray!10}
 $n_{Grant}$ & Number of granted independent claims \\
 
 $n_{App}$ & Number of applied independent claims \\
  \rowcolor{gray!10}
 $\Delta{Absolute}$ & $n_{App} - n_{Grant}$ \\
 
 $\Delta{Relative}$ & $\Delta{Absolute} / n_{App}$ \\
  \rowcolor{gray!10}
 $T_{grant} - T_{filing}$ & Time (in days) of grant date from filing date \\

Citation Count & Number of citations a granted patent has garnered in 8 years \\ 
 \rowcolor{gray!10}
 Any Maintenance Fee Paid & 1 if at least one of a 4, 8, or 12 Year Maintenance Fee has been paid for a granted patent, else 0 \\

 \end{tabular}
\end{table}

\subsection{Additional Patent-Specific Measures}

\textbf{Team size} denoted as $|(Inv)|$ is calculated according to Equation \eqref{eq:teamSize}, where $n_{total}$ is the number of members on an inventor team. Because team size is highly skewed, we transform it into a categorical variable to capture single-member team, small team, medium team, and large team dynamics.

\begin{equation}
|(Inv)| = \text{min}(n_{total}, 5)
\label{eq:teamSize}
\end{equation}

Patent applications can come from \textbf{big or small entities} (corporations), which is coded as a categorical variable where $e(p) = 1$ if a patent is from a big entity and $0$ otherwise. 
Inventor and examiner experience is calculated as described below.
\textbf{Examiner gender} is coded as a binary variable, where $exam(p) = 1$ if it is a female examiner and $0$ otherwise. 
We include fixed effects for \textbf{year} and \textbf{CPC code} at the class level. 
For the multiple country regression model, we have an additional categorical variable for \textbf{country}, as depicted in Table \ref{regAllCountry}.

\textbf{Field-level variation} in terms of gender differences in participation rates, thresholds to what is considered patentable or innovative, and differences in success rates could contribute to gender differences in success in innovation more broadly. 

These depict field-level gender differences in different parts of the patenting process from inventors to examiners, specifically, at the broader section level of CPC code for presentation ease, and variation was found to exist throughout several levels of the CPC hierarchy, for instance, persisting at the class level. 
In order to more adequately capture the effects of gender as opposed to effects stemming from field-level differences, we control for field as measured by focal \textbf{CPC code} at the class level.

In order to better isolate the effects of gender, we construct and include several salient features in our regression models.
The experience of the inventor teams and examiners involved could presumably affect the likelihood of a patent application being accepted \cite{frakes2015patent, frakes2017time}.

We measure the \textbf{average experience of the inventor team} to be the mean of the experience of all the inventor members.
Experience is measured by number of patent applications a member has been on at the application date considered. 
\textbf{Examiner experience} is measured by the number of applications an examiner has evaluated to date of the patent application in question. 

We validate our measure for \textbf{inventor experience} on a dataset consisting of granted USPTO patent applications from 1976 to present where inventor names are disambiguated. 
We construct a second measure, for granted applications, which consists of the average number of granted patents previous to the grant date of the focal application for the inventors on the team. 
We can only construct this second measure for inventors on granted patent applications due to only having access to disambiguated inventor data for granted patents. 
As a robustness check, we find that our measure of inventor experience as measured through average number of previous applications correlates highly with this second measure from disambiguated data where we measure the average number of previous granted applications, and this correlation is 0.841 using the Pearson correlation coefficient, lending validity to our measure used over the set of all applications.

\section{Unconventionality Computation}

\label{coOccurPatent}
We follow the literature in measuring novelty \cite{kim_technological_2016, youn_invention_2015} to calculate a measure to capture the level of unconventionality of a combination of categories. Each patent is assigned to a variety of technical areas that it pertains to, according to the “cooperative patent classification” (CPC) system. We utilize the tendency that inventions are categorized into multiple categories to create a CPC co-classification network in order to assess the unconventionality of a combination of categories. 

This is done by first creating a CPC co-occurrence network called $A_t$. $A_t$ is created cumulatively, covering the set of patents in the years at and below $t$. $A_t(i,j) = n$, where $n$ is the number of times categories $i$ and $j$ co-occur on a patent across all patents in the subset of time (of year at or below $t$) we are analyzing. 

To normalize for field-level variation in the tendencies of certain categories to be combined in particular ways by field, we adopt an approach developed in \cite{kim_technological_2016} to calculate the standard scores (z-scores) of CPC code pairs across the space of patents. 

\begin{equation}
z_{\alpha\beta} = \frac{o_{\alpha\beta} - \mu_{\alpha\beta }}{\sigma _{\alpha\beta}}
\label{eq:zScore}
\end{equation}

Equation \eqref{eq:zScore} represents the z-score of a pair of CPC codes $\alpha$ and $\beta$. 
The variable $o_{\alpha\beta}$ is the observed number of co-occurrences of $\alpha$ and $\beta$ within a patent in the data. Variables $\mu_{\alpha\beta }$ and $\sigma _{\alpha\beta}$ are the expected co-occurrences of the pair of CPC codes and its standard deviation, which is derived from a null model of the data in which CPC code arrangement is randomized while preserving CPC code usage and the number of patents in the subset of the data of years at or below year $t$.

A positive z-score would be associated with two CPC codes co-occurring more than expected, and a negative z-score would correspond with two CPC codes that are rarely combined relative to their expected co-occurrence \cite{kim_technological_2016}. 

In order to compute the expected co-occurrences of pairs of CPC codes, the baseline null model randomized the arrangement of CPC codes as in \cite{kim_technological_2016} to calculate the values of the variables in Equation \eqref{eq:zScore}.

We compute in a temporally evolving manner $A_t$ and $z_{\alpha\beta}$ where $\alpha,\beta \in \Phi$, where $\Phi$ represent the set of all CPC codes in the set of patents we analyze. 

This is done by cumulatively taking the set of all patents $P$, where $P(y) <= t$, where $y$ represents the year of the patent application. We look at applications in the range 2002 to 2018 to construct these z-score metrics, and the data spans years 2001 to 2018.
We calculate these measures based on the subclass level of the CPC code hierarchy. 

\subsection{Patent-level Unconventionality Construction}
\label{atypicalityPatent}
We have established how to compute the unconventionality of pairs of CPC codes being combined; however, we still need to compute a measure for unconventionality for a single patent. 
From Eq. \eqref{eq:zScore} we can compute the z-score measure for each pair of CPC codes across all CPC codes through time. 
This measure represents how the empirical combinations of areas differ from their expected combinations and gives a measure to represent whether a combination of CPC codes is unconventional or conventional. 
A negative z-score corresponds to an unconventional combination, while a positive z-score corresponds to a conventional combination. 
A z-score of $0$ would represent a neutral combination, neither unconventional nor conventional. 

For a given patent application $p$, say it combines three technological areas, or CPC codes $a$, $b$, and $c$, with $a$ being the focal, most relevant CPC code. 
We take all pairs of links with the focal code $a$, which in this example would be links $a,b$ and $a, c$, and take the minimum of their associated pairwise z-scores as representative of the most atypical, unconventional link: $min(z_{a,b}, z_{a,c})$. 
In more general terms, this can be written as follows:
\begin{equation}
L_{\textit{atypical}} = \min\limits_{i \in (1 \dots n)} z_{x_f, x_i}
\label{eq:atyp}
\end{equation}
In Eq. \eqref{eq:atyp}, $x_f$ represents the first, focal CPC code of a patent application, and $x_i$ where $i \in (1 \dots n)$ represents the $n$ other CPC codes in an application other than the focal CPC code. 
$L_{\textit{atypical}}$ thus represents the minimum pairwise distance between the focal CPC code and all other area codes in a particular patent application. 
A positive $L_{\textit{atypical}}$ corresponds to a high minimum z-score, representing a conventional combination of CPC codes, while a negative $L_{\textit{atypical}}$ corresponds to a low minimum z-score, representing an atypical combination of CPC codes and thus an atypical, unconventional patent application. 

We can then negate $L_{\textit{atypical}}$ as in Eq. \eqref{eq:atypInv}. 
\begin{equation}
p_{atypical} = (-1) \cdot L_{\textit{atypical}} 
\label{eq:atypInv}
\end{equation}
 
Thus, $p_{atypical}$ represents the level of unconventionality of patent $p$, and the scale is inverted: a positive value for $p_{atypical}$ represents an unconventional patent (the more positive, the more atypical and unconventional), and a negative value for $p_{atypical}$ represents a conventional patent (and larger negative values correspond to more deeply conventional patents).

\addcontentsline{toc}{section}{Findings}
\part*{Findings}

\section{Robustness with Several Measures of Unconventionality Innovation}
\label{severalAtyp}
The main measure for unconventionality we use is $L_{\textit{atypical}}$ from Eq. \eqref{eq:atyp}. 
As described previously, $L_{\textit{atypical}}$ is the minimum pairwise distance between the focal CPC code and all other area codes in a particular patent application. 
Additional measures of unconventionality include the mean and median of the pairwise distances from the focal CPC code to all other codes in an application \cite{kim_technological_2016}, and the measure based on the mean is depicted in Eq. \eqref{eq:atypMean} and the measure based on the median is depicted in Eq. \eqref{eq:atypMed}.

\begin{equation}
L_{\textit{atypicalMean}} =\overline{z_{x_f, x_i}} \textit{ where } i \in (1 \dots n) 
\label{eq:atypMean}
\end{equation}

\begin{equation}
L_{\textit{atypicalMedian}} = med({z_{x_f, x_i}}) \textit{ where } i \in (1 \dots n) 
\label{eq:atypMed}
\end{equation}

We construct patent level unconventionality in the same fashion as Eq. \eqref{eq:atypInv} for the measures based on the mean and median. 
We can negate $L_{\textit{atypicalMean}}$ and $L_{\textit{atypicalMedian}}$ as in Equations \eqref{eq:atypInvMean} and \eqref{eq:atypInvMed} to create these additional measures for unconventionality.

\begin{equation}
p_{atypicalMean} = (-1) \cdot L_{\textit{atypicalMean}} 
\label{eq:atypInvMean}
\end{equation}

\begin{equation}
p_{atypicalMedian} = (-1) \cdot L_{\textit{atypicalMedian}} 
\label{eq:atypInvMed}
\end{equation}

\begin{table}[!htbp]
 \caption{\textbf{Patent acceptance drops for women and rise for men at all levels of unconventionality.} Regressions results for all unconventionality measures: $p_{atypical}$ (as denoted by `Measure 1'), $p_{atypicalMean}$ (as denoted by `Measure 2'), and $p_{atypicalMedian}$ (as denoted by `Measure 3'). There are fixed effects for year, team size, and CPC Code. Model 1 is in the main text, models 2 and 3 are confirmatory models. This table corresponds to the interaction effect depicted in Figure 1B (main text). The operationalization of the measures is described in Section \ref{severalAtyp}. The penalty for unconventional innovation for women increases with the percentage of women on the team at any level of unconventionality.} 
 \label{mainAtypicalAll}
 \begin{center}

 \resizebox{1\width}{!}{


\def\sym#1{\ifmmode^{#1}\else\(^{#1}\)\fi}
\begin{tabularx}{\linewidth}{ p{\dimexpr.29\linewidth-2\tabcolsep-1.3333\arrayrulewidth} X X X }  
\toprule
DV = Patent Acceptance & Measure 1 & Measure 2 & Measure 3 \\
\midrule
Women Majority \\Inventors &  -0.211\sym{***} & -0.222\sym{***} & -0.221\sym{***} \\
                            &    (7.66e-3)    &    (8.03e-3)    &    (7.98e-3)   \\
\addlinespace
Unconventionality &     3.08e-4\sym{***} & 3.27e-4\sym{***} &   3.29e-4\sym{***} \\
            & (1.29e-5)            & (1.28e-5)            & (1.28e-5)              \\
\addlinespace
Women Majority \\Inventors $\times$ Unconventionality &   -1.42e-4\sym{***} & -1.89-4\sym{***} & -1.83e-4\sym{***} \\
                                                 &  (2.82e-5)            &  (2.68e-5)            &  (2.69e-5)                \\
\addlinespace
Big Entity &    0.682\sym{***} &    0.681\sym{***}  &    0.682\sym{***} \\
           & (4.58e-3)            & (4.58e-3)            & (4.58e-3)              \\
\addlinespace
Female Examiner &      -0.123\sym{***} &      -0.123\sym{***} &      -0.123\sym{***}\\
                 & (4.50e-3)            & (4.50e-3)            & (4.50e-3)                   \\
\addlinespace
Average Inventor \\Experience &     -3.27e-4\sym{***} &     -3.23e-4\sym{***}  &     -3.20e-4\sym{***} \\
                            & (3.10e-5)                    & (3.10e-5)                    & (3.10e-5)                           \\
\addlinespace
Average Examiner \\Experience &     1.29e-3\sym{***}  &     1.29e-3\sym{***} &    1.29e-3\sym{***}\\
                            & (7.37e-6)                  & (7.37e-6)                  & (7.37e-6)                      \\
\addlinespace
Year FE &     Y  &     Y &    Y\\
\addlinespace
Team Size FE &     Y  &     Y &    Y\\
\addlinespace
CPC Code FE &     Y  &     Y &    Y\\
\addlinespace
\midrule
Observations        &     1,299,721    & 1,299,721 & 1,299,721 \\
Pseudo \(R^{2}\)    &        0.0898 &         0.0898 &        0.0898     \\
\bottomrule
\multicolumn{4}{l}{\footnotesize The base case for the indicator of Women Majority Inventors is Men Majority Inventors.}\\
\multicolumn{4}{l}{\footnotesize Standard errors in parentheses}\\
\multicolumn{4}{l}{\footnotesize \sym{*} \(p<0.05\), \sym{**} \(p<0.01\), \sym{***} \(p<0.001\)}\\
\end{tabularx}
}
	 
\end{center}

\end{table}

\begin{table}[!htbp]
 \caption{\textbf{Model robust to iterative building.} Regression results for iteratively built logistic regression models. Including fixed effects for year, team size, and primary CPC Code. From left to right column-wise, each logistic regression model in the table builds to include the features in the columns to the left of it - Model 1: Controls Only, Model 2: With Examiner Features, Model 3: With Inventor Features (no gender), Model 4: With Inventor Gender Feature, Model 5: With Unconventionality, Model 6: With Gender/Unconventionality Interaction} 
 \label{iterativeMainUSARegRestuls}
 \begin{center}

 \resizebox{1\width}{!}{

\resizebox{0.95\textwidth}{!}{%
\def\sym#1{\ifmmode^{#1}\else\(^{#1}\)\fi}
\small
\renewcommand{\arraystretch}{0.8}

\begin{tabularx}{\linewidth}{ p{\dimexpr.15\linewidth-2\tabcolsep-1.3333\arrayrulewidth} X X X X X X X}

\toprule
DV = Patent Acceptance &  Controls Only & Examiner Features & Inventor Features (no gender) & Inventor Gender Feature & Unconv.$^1$ & Gender $\times$ Unconv. \\
\midrule
Women \\Majority \\Inventors &  --- & --- &  --- & -0.196\sym{***} & -0.191\sym{***} & -0.211\sym{***} \\
                            &        &    & & (6.59e-3)&   (6.60e-3)  &    (7.66e-3)   \\
\addlinespace
Unconv. &     --- & --- &     --- & --- & 2.85e-4\sym{***} &3.08e-4\sym{***} \\
            &              &         &&& (1.21e-5)  & (1.29e-5)           \\
\addlinespace
Women \\Majority \\Inventors $\times$ \\Unconv. &   --- & ---  &  --- & --- &--- & -1.42e-4\sym{***} \\
                           &           &&&     &             &  (2.82e-5)                \\
\addlinespace
Big Entity &    ---&  --- &  0.682\sym{***} & 0.680\sym{***} &     0.682\sym{***}  &     0.682\sym{***} \\
           &              &    & (4.49e-3) &(4.57e-3) &  (4.58e-3)    & (4.58e-3)             \\
\addlinespace
Female\\Examiner &   ---    &  -0.132\sym{***} & -0.127\sym{***} & -0.127\sym{***} &     -0.123\sym{***} &      -0.123\sym{***}\\
                 &           &   (4.38e-3) &(4.42e-3)& (4.49e-3) & (4.50e-3)     & (4.50e-3)                    \\
\addlinespace
Average \\Inventor \\Experience &     --- &  --- & -3.24e-4\sym{***} & -3.37e-4\sym{***} &   -3.30e-4\sym{***}  &     -3.27e-4\sym{***} \\
                            &          & & (3.07e-5) & (3.11e-5)   &   (3.10e-5) & (3.10e-5)        \\
\addlinespace
Average \\Examiner \\Experience &     --- &  1.27e-3\sym{***} & 1.28e-3\sym{***} & 1.28e-3\sym{***} &    1.29e-3\sym{***}  &     1.29e-3\sym{***}\\
                            &            & (7.17e-6) & (7.23e-6) &    (7.35e-6)     &      (7.37e-6)     & (7.37e-6)                       \\
\addlinespace
Year FE &     Y  &     Y &    Y &    Y &    Y &    Y\\
\addlinespace
Team \\Size FE &     Y  &     Y &    Y & Y &    Y &    Y\\
\addlinespace
CPC \\Code FE &     Y  &     Y &    Y &  Y &    Y &    Y\\
\midrule
Observations        &      1,340,366     &  1,338,831  & 1,338,824 &     1,299,721 &  1,299,721 &   1,299,721   \\
Pseudo \(R^{2}\)    &       0.0533 &       0.0751 &        0.0888   &         0.0894 &       0.0897   &         0.0898   \\
AIC    &          1593954 &         1554075   &       1531093  &       1484396  &         1483845     &        1483821    \\
\bottomrule
\multicolumn{7}{l}{\footnotesize The base case for the indicator of Women Majority Inventors is Men Majority Inventors.}\\
\multicolumn{7}{l}{\footnotesize $^1$ Unconv. stands for unconventionality.}\\
\multicolumn{7}{l}{\footnotesize Standard errors in parentheses}\\
\multicolumn{7}{l}{\footnotesize \sym{*} \(p<0.05\), \sym{**} \(p<0.01\), \sym{***} \(p<0.001\)}\\
\end{tabularx}
}}
	 
\end{center}

\end{table}

\section{Gender Penalty and Robustness to Several Gender Composition Operationalizations}

One metric for inventor team gender composition we use is majority women or men inventors, as described previously. The regression results using this metric are depicted in Table \ref{mainAtypicalAll} in the column depicting $p_{atypical}$. 
We find that not only is there a negative coefficient for women majority inventors, but that the interaction between women majority inventors and unconventionality is negative.

To evaluate model robustness, we build the logistic regression model up iteratively and see that adding the features in the full model results in the lowest AIC score, as depicted in Table \ref{iterativeMainUSARegRestuls}.

As further robustness checks, we also use alternate measures of inventor team gender composition. These include a gender operationalization indicating whether a team is majority female, majority male, or 50\% Women and 50\% Men, as well as a gender operationalization indicating whether the inventor team is all female or not. The regression results for both of these additional gender metrics are depicted in Tables \ref{regFirstAuth} and \ref{regMajority}.

For the team gender composition of male or female majority, or 50\% Men-50\% Women teams regressions results in Table \ref{regFirstAuth}, we see patent applications with women majority teams face a penalty in terms of patent grant probability as compared to the men majority base rate. Teams of 50\% Men-50\% Women also face a penalty compared to men majority teams, and less of a penalty than women majority teams, suggesting having more men reduces this penalty. We observe that unconventionality has a positive relationship with grant probability, while the interaction between women majority and 50\% Men-50\% Women teams is negative, more strongly so when it is a women majority team. Note here that the base case is for male majority teams for both the inventor gender main effect as well as the interaction terms. 

Similarly, for the all-women inventor team regressions results in Table \ref{regMajority}, we see all-women teams face a penalty in terms of predicting patent acceptance. We observe unconventionality with a positive effect as well as the effect of all-women teams interacted with unconventionality is negative. These robustness checks yield results very similar to the gender metric of the women versus male majority inventor teams discussed in the main text.

We also observe for the continuous variable of the proportion of women inventors gender operationalization regressions results in Table \ref{regContFem}, we see that teams with a greater proportion of women inventors face a penalty in terms of predicting patent acceptance. We observe unconventionality with a positive effect and that the effect of proportion of women inventors interacted with unconventionality is negative. These robustness checks also yield confirmatory results with the gender metric discussed in the main text. 

\begin{table}[!htbp]
 \caption{\textbf{Gender penalty results robust to different gender operationalization of team gender composition.} Regressions results for gender composition of inventor team variable operationalization with women majority, men majority, and 50\% Men-50\% Women teams. There are controls and fixed effects for year, team size, and CPC Code.} \label{regFirstAuth}
 \begin{center}

 \resizebox{1\width}{!}{{
\def\sym#1{\ifmmode^{#1}\else\(^{#1}\)\fi}
\begin{tabular}{l*{1}{c}}
\toprule
         DV = Patent Acceptance            &\multicolumn{1}{c}{(US Patent Applications)}         \\
\midrule
              &                     \\
Women Majority Team           &        -0.257\sym{***}\\
                    &    (1.06e-2)         \\
\addlinespace

50\% Men-50\% Women Team           &      -0.166\sym{***}\\
                    &    (1.06e-2)         \\
\addlinespace

Unconventionality            &    3.08e-4\sym{***}\\
                    & (1.29e-5)         \\
\addlinespace
Women Majority Team $\times$ Unconventionality&   -2.38e-4\sym{***}\\
                    &  (3.87e-5)         \\
\addlinespace

50\% Men-50\% Women Team  $\times$ Unconventionality&   -5.28e-5\\
                    &  (3.78e-5)         \\
\addlinespace

Big Entity             &    0.682\sym{***}\\
                    & (4.58e-3)         \\
\addlinespace

\addlinespace
Female Examiner            &     -0.123\sym{***}\\
                    & (4.50e-3)         \\
\addlinespace

\addlinespace
Average Inventor Experience           &    -3.29e-4\sym{***}\\
                    & (3.11e-5)         \\
\addlinespace

\addlinespace
Average Examiner Experience           &    1.29e-3\sym{***}\\
                    & (7.37e-6)         \\
\addlinespace
Year FE          &     Y \\
\addlinespace
Team Size FE          &     Y \\
\addlinespace
CPC Code FE          &     Y \\
\addlinespace

\midrule
Observations        &       1,299,721      \\
Pseudo \(R^{2}\)    &         0.0898         \\
\bottomrule
\multicolumn{2}{l}{\footnotesize The base case for inventor team gender is men majority team.}\\ 
\multicolumn{2}{l}{\footnotesize The base case for big entity is small entity.}\\
\multicolumn{2}{l}{\footnotesize Standard errors in parentheses}\\
\multicolumn{2}{l}{\footnotesize \sym{*} \(p<0.05\), \sym{**} \(p<0.01\), \sym{***} \(p<0.001\)}\\
\end{tabular}
}
}

\end{center}

\end{table}

\begin{table}[!htbp]
 \caption{\textbf{Gender penalty results robust to different gender operationalization of all-women inventor teams.} Regressions results for all-women inventor team indicator variable operationalization. There are controls and fixed effects for year, team size, and CPC Code.} \label{regMajority}
 \begin{center}

 \resizebox{1\width}{!}{{
\def\sym#1{\ifmmode^{#1}\else\(^{#1}\)\fi}
\begin{tabular}{l*{1}{c}}
\toprule
        DV = Patent Acceptance              &\multicolumn{1}{c}{(US Patent Applications)}         \\
\midrule
             &                     \\
All-Women Inventor Team           &       -0.278\sym{***}\\
                    &    (1.33e-2)         \\
\addlinespace

Unconventionality              &     2.94e-4\sym{***}\\
                    & (1.21e-5)         \\
\addlinespace
All-Women Inventor Team  $\times$ Unconventionality&   -2.79e-4\sym{***}\\
                    &  (5.28e-5)         \\
\addlinespace

Big Entity             &    0.683\sym{***}\\
                    & (4.50e-3)         \\
\addlinespace

\addlinespace
Female Examiner            &      -0.123\sym{***}\\
                    & (4.43e-3)         \\
\addlinespace

\addlinespace
Average Inventor Experience           &     -3.14e-4\sym{***}\\
                    & (3.08e-5)         \\
\addlinespace

\addlinespace
Average Examiner Experience           &     1.29e-3\sym{***}\\
                    & (7.25-6)         \\
\addlinespace
Year FE          &     Y \\
\addlinespace
Team Size FE          &     Y \\
\addlinespace
CPC Code FE          &     Y \\
\addlinespace

\midrule
Observations        &       1,338,824 \\
Pseudo \(R^{2}\)    &        0.0894         \\
\bottomrule
\multicolumn{2}{l}{\footnotesize The base case for big entity is small entity.}\\
\multicolumn{2}{l}{\footnotesize Standard errors in parentheses}\\
\multicolumn{2}{l}{\footnotesize \sym{*} \(p<0.05\), \sym{**} \(p<0.01\), \sym{***} \(p<0.001\)}\\
\end{tabular}
}
}

\end{center}

\end{table}

\begin{table}[!htbp]
 \caption{\textbf{Gender penalty results robust to different gender operationalization of all-women inventor teams.} Regressions results for proportion of women inventors variable operationalization. There are controls and fixed effects for year, team size, and CPC Code.} \label{regContFem}
 \begin{center}

 \resizebox{1\width}{!}{{
\def\sym#1{\ifmmode^{#1}\else\(^{#1}\)\fi}
\begin{tabular}{l*{1}{c}}
\toprule
        DV = Patent Acceptance              &\multicolumn{1}{c}{(US Patent Applications)}         \\
\midrule
             &                     \\
Proportion of Women Inventors          &       -0.309\sym{***}\\
                    &    (1.01e-2)         \\
\addlinespace

Unconventionality              &     2.90e-4\sym{***}\\
                    & (1.37e-5)         \\
\addlinespace
Proportion of Women Inventors  $\times$ Unconventionality&   -1.09e-4\sym{**}\\
                    &  (3.99e-5)         \\
\addlinespace

Big Entity             &    0.681\sym{***}\\
                    & (4.50e-3)         \\
\addlinespace

\addlinespace
Female Examiner            &      -0.121\sym{***}\\
                    & (4.43e-3)         \\
\addlinespace

\addlinespace
Average Inventor Experience           &     -3.04e-4\sym{***}\\
                    & (3.08e-5)         \\
\addlinespace

\addlinespace
Average Examiner Experience           &     1.29e-3\sym{***}\\
                    & (7.25-6)         \\
\addlinespace
Year FE          &     Y \\
\addlinespace
Team Size FE          &     Y \\
\addlinespace
CPC Code FE          &     Y \\
\addlinespace

\midrule
Observations        &       1,338,824 \\
Pseudo \(R^{2}\)    &        0.0898        \\
\bottomrule
\multicolumn{2}{l}{\footnotesize The base case for big entity is small entity.}\\
\multicolumn{2}{l}{\footnotesize Standard errors in parentheses}\\
\multicolumn{2}{l}{\footnotesize \sym{*} \(p<0.05\), \sym{**} \(p<0.01\), \sym{***} \(p<0.001\)}\\
\end{tabular}
}
}

\end{center}

\end{table}

\begin{figure}[!ht]
 \centering
 \includegraphics[width=0.6\linewidth]{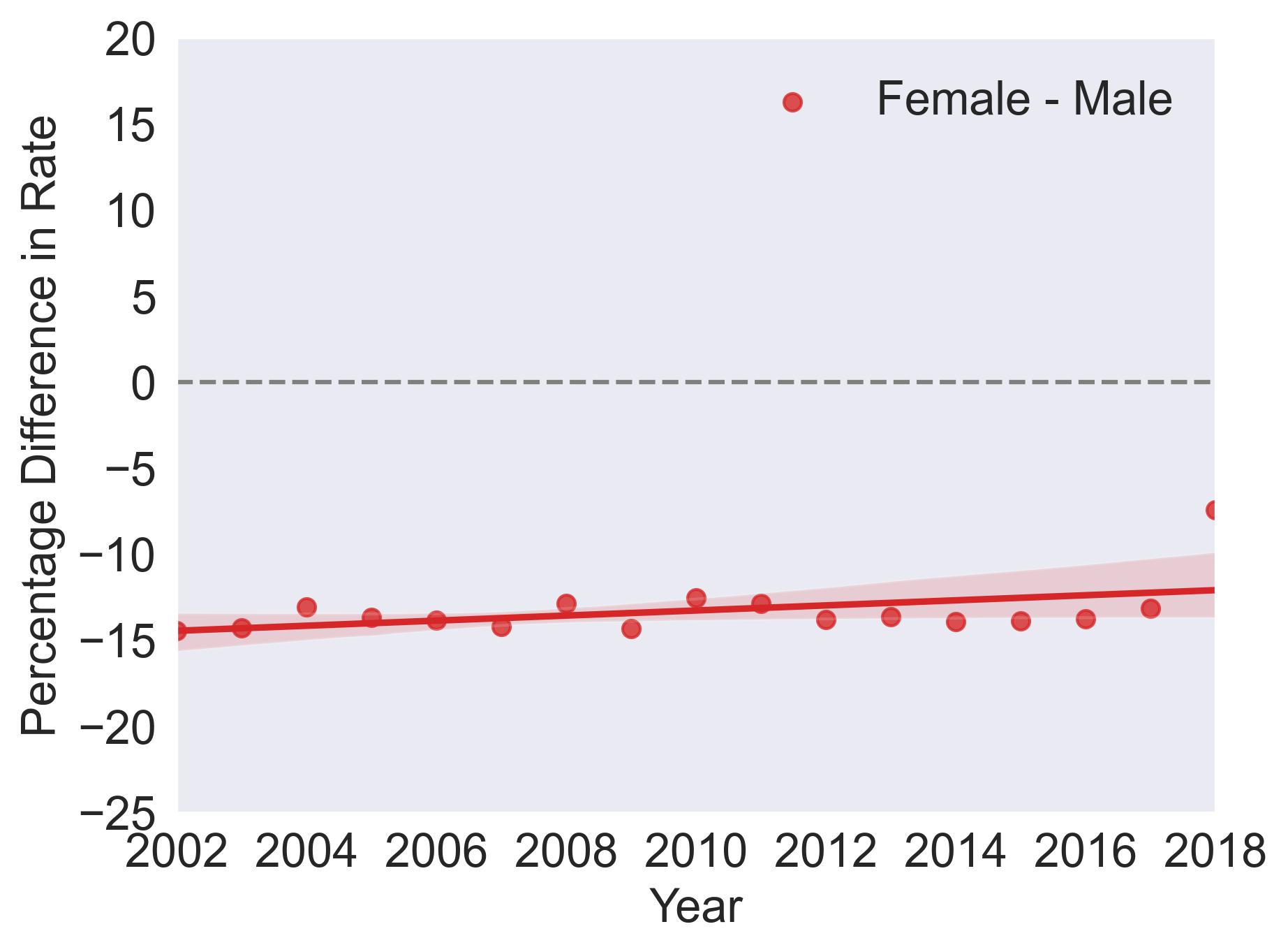}
 \caption{\textbf{Women are less likely to be awarded a patent than are men through time.} The data represents 4,754,198 patents submitted to the U.S. Patent Office (USTPO) between 2002 and 2018. The gender difference (between single-gender teams) in success in being awarded a patent is persistent across time despite significant increases in the number of women innovators seeking patents and the nearly 1-to-1 ratio of women to men in STEM college majors responsible for patenting including chemistry, biology, math, and other fields \cite{cimpian2020understanding}. }
 \label{fig:genderFig1}
\end{figure}

Figure \ref{fig:genderFig1} shows that in our data, we observe the well-documented finding that women are significantly less likely to be awarded a patent than men. 

Figure \ref{fig:genderFig1} indicates that the relatively lower success rate for women has been essentially constant over the last two decades despite the increase in women’s participation in patenting, and the near 1-to-1 ratio in college majors including chemistry, biology, math, and other fields relevant to patenting \cite{cimpian2020understanding}. In 2018, women were 7.4 percentage points less likely than men to be granted a patent (chi-squared test p $<$ 0.001), which equates to men having about 9.1\% better odds of being granted a patent.

\section{Reinforcing Confirmatory Regression Analysis on Reduction of Claims and Time Delays for Women Innovators}

 A critical part of the patent review process focuses on a patent’s legal claims. Claims give the owner the exclusive legal right to use the patented innovation or license its use to others. Thus, the more claims denied, the less control an innovator has over their innovation’s use and future financial returns. Nevertheless, claims rely on the examiner’s subjective standards about patent-eligibility \cite{medeiros2016fintech}. 
 
Each patent must have at least one independent claim delineating the legal rights to the intellectual property claimed by the applicant. Ultimately, these claims are the enforceable elements of a granted patent.

We analyze changes to the number of independent claims from the time of application to granted versions of the patent text. 
Let $n_{App} = |\text{Independent Claims at Application Stage}|$.
Let $n_{Grant} = |\text{Independent Claims at Granted Stage}|$. 
The features we focus on are denoted in Equations \eqref{eq:absoluteDelta} and \eqref{eq:relativeClaims} representing the difference in number of independent claims at application stage minus grant stage as well as the relative difference in number of claims as compared to the application stage respectively. This reflects the changes to the number of independent claims requested by the applicant and those granted by the patent office.

\begin{equation}
\Delta{Absolute} = n_{App} - n_{Grant}
\label{eq:absoluteDelta}
\end{equation}

\begin{equation}
\Delta{Relative} = \frac{n_{App} - n_{Grant}}{n_{App}} = \frac{\Delta{Absolute}}{n_{App}}
\label{eq:relativeClaims}
\end{equation}

Regressions on granted patents to predict the quantities regarding the absolute and relative claims at time of filing versus at time are grant, namely metrics $\Delta{Absolute}$ and $\Delta{Relative}$ respectively, are denoted in first two columns of Table \ref{mergedClaims}. 
We see that being a women majority team increases $\Delta{Absolute}$ and $\Delta{Relative}$. 
This suggests women inventors face a reduction in the number of claims they are granted from those they request on their patent applications, suggesting increased claim denials than male inventors.

\begin{table}[!htbp]
 \caption{\textbf{Women experience higher denials in claims than men innovators.} $\Delta{Absolute}$, $\Delta{Relative}$, and $n_{grant}$ regression results depicted. There are controls for entity size, examiner gender, inventor experience, examiner experience, and fixed effects for year, team size, and CPC code.} \label{mergedClaims}
 \begin{center}
 
 \resizebox{1\width}{!}{{
\def\sym#1{\ifmmode^{#1}\else\(^{#1}\)\fi}
\begin{tabular}{l*{3}{c}}
\toprule
            DV        &\multicolumn{1}{c}{$\Delta{Absolute}$}   &\multicolumn{1}{c}{$\Delta{Relative}$}   &\multicolumn{1}{c}{$n_{grant}$}         \\
\midrule
           &                     \\
Women Majority Inventors           &       0.108\sym{***} & 3.26e-2\sym{***} & -7.54e-2\sym{***}\\
                    &    (1.37e-2)    & (4.70e-3) & (1.04e-2)   \\
\addlinespace
Unconventionality             &    4.32e-4\sym{***} & 8.49e-5\sym{***}  & 6.87e-5\sym{***}  \\
                    & (2.55e-5)   & (8.71e-6) & (1.92e-5)       \\

\addlinespace

Big Entity             &    -0.320\sym{***} & -9.66e-2\sym{***}  & 0.156\sym{***}  \\
                    & (9.68e-3)   & (3.31e-3) & (7.30e-3)       \\              

\addlinespace

Female Examiner            &    3.27e-2\sym{***} & 7.30e-3\sym{*}  & -3.49e-2\sym{***}  \\
                    & (8.97e-3)   & (3.07e-3) & (6.76e-3)       \\
                    
\addlinespace

Average Inventor Experience         &     1.97e-3\sym{***} & 2.43e-4\sym{***} & -3.62e-4\sym{***} \\
     & (6.79e-5)   & (2.32e-5) & (5.12e-5)       \\
\addlinespace

Examiner Experience             &    -1.48e-4\sym{***} & -5.29e-5\sym{***}  & 5.90e-5\sym{***}  \\
                    & (1.19e-5)   & (4.06e-6) & (8.96e-6)       \\

\addlinespace

Year FE          &     Y & Y & Y \\
\addlinespace
Team Size FE          &     Y& Y & Y  \\
\addlinespace
CPC Code FE          &     Y& Y & Y\\
\addlinespace
\midrule
Observations        &      830,567    & 830,567 &  830,567   \\
\(R^{2}\)    &        0.0497 & 0.0749 &  0.1348   \\
\bottomrule
\multicolumn{3}{l}{\footnotesize The base case for the indicator of Women Majority Inventors is Men Majority Inventors.}\\
\multicolumn{2}{l}{\footnotesize Standard errors in parentheses}\\
\multicolumn{2}{l}{\footnotesize \sym{*} \(p<0.05\), \sym{**} \(p<0.01\), \sym{***} \(p<0.001\)}\\
\end{tabular}
}
}

\end{center}

\end{table}

We also find that when predicting the number of granted independent claims $n_{grant}$, being a women majority inventor team negatively predicts the number of granted independent claims, while a higher unconventionality level positively predicts this measure $n_{grant}$. This is depicted in the last column of Table \ref{mergedClaims}.

We measure ``patent review latency'' or the time between the patent application filing date and the final patentability decision. The greater the latancy, the more the innovation’s diffusion in the marketplace and the innovator’s financial returns are delayed \cite{ahmadpoor2017dual}. 
Regression results on granted patent applications to predict the quantity ${T_{grant} - T_{filing}}$, which represents the time (in days) of grant date from filing date, are denoted in Table \ref{timeDelta}. 
We see that being a women majority inventor team increases ${T_{grant} - T_{filing}}$. 
This suggests that women inventors have to wait longer for patent protection than their male counterparts. 
We also observe that increased unconventionality increases ${T_{grant} - T_{filing}}$. 
This indicates that unconventional applications are more likely to take longer to be granted.

\begin{table}[!htbp]
 \caption{\textbf{Women face more delays than men in grant time for patent applications.} Regression Results for time (in days) of grant date from filing date, including controls for inventor experience and fixed effects for year, team size, and CPC code.} \label{timeDelta}
 \begin{center}
 
 \resizebox{1\width}{!}{{
\def\sym#1{\ifmmode^{#1}\else\(^{#1}\)\fi}
\begin{tabular}{l*{1}{c}}
\toprule
            DV = ${T_{grant} - T_{filing}}$           &\multicolumn{1}{c}{(USA Granted Patents)}         \\
\midrule
          &                     \\
Women Majority Inventors           &        18.180\sym{***}\\
                    &    (1.966)         \\
\addlinespace
Unconventionality              &    0.209\sym{***}\\
                    & (3.645e-3)         \\
\addlinespace

Big Entity             &    4.586\sym{***}\\
                    & (1.385)         \\
\addlinespace

Female Examiner              &    18.236\sym{***}\\
                    & (1.283)         \\
\addlinespace

Average Inventor Experience         &     -0.295\sym{***} \\
     & (9.719e-3)        \\
\addlinespace

Examiner Experience         &     -0.279\sym{***} \\
     & (1.701e-3)        \\
\addlinespace

Year FE          &     Y \\
\addlinespace
Team Size FE          &     Y \\
\addlinespace
CPC Code FE          &     Y \\
\addlinespace
\midrule
Observations        &      830,553       \\
\(R^{2}\)    &         0.3315      \\
\bottomrule
\multicolumn{2}{l}{\footnotesize The base case for the indicator of Women Majority Inventors is Men Majority Inventors.}\\
\multicolumn{2}{l}{\footnotesize Standard errors in parentheses}\\
\multicolumn{2}{l}{\footnotesize \sym{*} \(p<0.05\), \sym{**} \(p<0.01\), \sym{***} \(p<0.001\)}\\
\end{tabular}
}
}

\end{center}

\end{table}

\section{Examination Experience and the Gender Innovation Penalty}

We observe a gender penalty for women when they engage in unconventional innovation.
One possible explanation is that examiners have difficulty assessing the quality of truly unconventional inventions if they do not have as much exposure to unconventional innovations.
The tendency to penalize unconventionality might be particularly strong for less experienced examiners and this tendency to penalize unconventional innovations might go down with more examination experience. 
We observe that the unconventionality grant rate goes up as examiners gain more experience as depicted in the regression results of Table \ref{examinerExpRegMain}. 
This is specifically observed in the interaction effect between examiner experience level and the level of unconventionality of the innovation, even while the main effect for unconventionality is not significant in this regression the interaction is significant. 
This suggests that perhaps more experienced examiners are more open to highly unusual, unconventional inventions.

\begin{table}[!htbp]
 \caption{\textbf{Higher examiner experience reduces tendency to penalize unconventional innovations.} Logistic regressions results with controls and fixed effects for year, team size, and CPC Code.} \label{examinerExpRegMain}
 \begin{center}

 \resizebox{1\width}{!}{{
\resizebox{0.89\textwidth}{!}{%
\def\sym#1{\ifmmode^{#1}\else\(^{#1}\)\fi}
\begin{tabular}{l*{1}{c}}
\toprule
         DV = Patent Acceptance            &\multicolumn{1}{c}{(US Patent Applications)}         \\
\midrule
              &                     \\
Women Majority Inventors           &        -0.211\sym{***}\\
                    &    (7.67e-3)         \\
\addlinespace

Unconventionality             &    -3.47e-5\\
                    & (2.66e-5)         \\
\addlinespace
Examiner Experience Level Mid &   0.532\sym{***}\\
                    &  (5.90e-3)         \\
\addlinespace
Examiner Experience Level High &   1.08\sym{***}\\
                    &  (6.39e-3)         \\
\addlinespace
Women Majority Inventors $\times$ Unconventionality&   -1.28e-4\sym{***}\\
                    &  (2.80e-5)         \\
\addlinespace

Examiner Experience Level Mid $\times$ Unconventionality&   3.09e-4\sym{***}\\
                    &  (3.05e-5)         \\
\addlinespace
Examiner Experience Level High $\times$ Unconventionality&   5.35e-4\sym{***}\\
                    &  (2.93e-5)         \\
\addlinespace

Big Entity             &    0.685\sym{***}\\
                    & (4.58e-3)         \\
\addlinespace

\addlinespace
Female Examiner            &     -0.141\sym{***}\\
                    & (4.50e-3)         \\
\addlinespace

\addlinespace
Average Inventor Experience           &    -3.68e-4\sym{***}\\
                    & (3.11e-5)         \\
\addlinespace
Year FE          &     Y \\
\addlinespace
Team Size FE          &     Y \\
\addlinespace
CPC Code FE          &     Y \\
\addlinespace

\midrule
Observations        &      1,299,721       \\
Pseudo \(R^{2}\)    &        0.0892 \\
\bottomrule
\multicolumn{2}{l}{\footnotesize The base case for the indicator of Women Majority Inventors is Men Majority Inventors.}\\
\multicolumn{2}{l}{\footnotesize The base case for examiner experience level is low-level experience.}\\
\multicolumn{2}{l}{\footnotesize The base case for big entity is small entity.}\\
\multicolumn{2}{l}{\footnotesize Standard errors in parentheses}\\
\multicolumn{2}{l}{\footnotesize \sym{*} \(p<0.05\), \sym{**} \(p<0.01\), \sym{***} \(p<0.001\)}\\
\end{tabular}
}}
}

\end{center}

\end{table}

Furthermore, we find that for unconventional patent applications, being a women majority inventor team is negatively predictive of examiner experience. 
This is depicted in the linear regression results of Table \ref{examinerExperienceFemaleInvReg}. 
This negative coefficient suggests that women inventors are getting less experienced examiners evaluating their patent applications for unconventional work, exacerbating the potential for penalizing that less experienced examiners may rely on in evaluating unconventional inventions and thus contributing to the lower grant rates women inventors experience when engaging in unconventional work.

\begin{table}[!htbp]
 \caption{\textbf{Women inventors that are unconventional are assigned less experienced examiners than men on average.} Linear regression results with fixed effects for year, team size, and CPC code done on the set of patent applications that are unconventional.} \label{examinerExperienceFemaleInvReg}
 \begin{center}

 \resizebox{1\width}{!}{


{
\def\sym#1{\ifmmode^{#1}\else\(^{#1}\)\fi}
\begin{tabular}{l*{1}{c}}
\toprule
         DV = Examiner Experience            &\multicolumn{1}{c}{(Unconventional US Patent Applications)}         \\
\midrule
              &                     \\
Women Majority Inventors           &        -5.76\sym{**}\\
                    &    (1.82)         \\

\addlinespace
Year FE          &     Y \\
\addlinespace
Team Size FE          &     Y \\
\addlinespace
CPC Code FE          &     Y \\
\addlinespace

\midrule
Observations        &     405,391       \\
\(R^{2}\)    &        0.2862 \\
\bottomrule
\multicolumn{2}{l}{\footnotesize The base case for the indicator of Women Majority Inventors is Men Majority Inventors.}\\
\multicolumn{2}{l}{\footnotesize Standard errors in parentheses}\\
\multicolumn{2}{l}{\footnotesize \sym{*} \(p<0.05\), \sym{**} \(p<0.01\), \sym{***} \(p<0.001\)}\\
\end{tabular}
}
}

\end{center}

\end{table}

\section{Homophily Between Women Inventors and Woman Examiners Exacerbates Gender Penalty}

We observe gender homophily between women inventors and women examiners in the USPTO patent applications. 

Women inventors are over-represented in some areas of inventing, and likewise, women examiners are over-represented in some areas of patent examination. 
We find that being an all-female inventor team is positively predictive of assignment to a female examiner, as depicted in the regression results of Table \ref{femHomophily}. 
\begin{table}[!h]
 \caption{\textbf{Women homophily between examiners and inventors exacerbates systemic gender inequalities.} Logistic regression results with fixed effects for year, team size, and CPC code. This regression supports the bivariate data presented in Figure 2C in the main text. } \label{femHomophily}
 \begin{center}

 \resizebox{1\width}{!}{{
\def\sym#1{\ifmmode^{#1}\else\(^{#1}\)\fi}
\begin{tabular}{l*{1}{c}}
\toprule
         DV = Female Examiner            &\multicolumn{1}{c}{(US Patent Applications)}         \\
\midrule
              &                     \\

All-Female Inventor Team          &        0.156\sym{***}\\
                    &    (1.22e-2)         \\

\addlinespace
Year FE          &     Y \\
\addlinespace
Team Size FE          &     Y \\
\addlinespace
CPC Code FE          &     Y \\
\addlinespace

\midrule
Observations        &       940,751        \\
Psuedo \(R^{2}\)    &         0.0437      \\
\bottomrule
\multicolumn{2}{l}{\footnotesize On the subset of applications that have either a man or woman examiner and }\\
\multicolumn{2}{l}{\footnotesize either an all-men or all-women inventor team.}\\
\multicolumn{2}{l}{\footnotesize Standard errors in parentheses}\\
\multicolumn{2}{l}{\footnotesize \sym{*} \(p<0.05\), \sym{**} \(p<0.01\), \sym{***} \(p<0.001\)}\\
\end{tabular}
}
}

\end{center}

\end{table}
This regression is done between all-women and all-men inventor teams on applications with either men or women examiners.
We find that all-women inventor team patent applications are 16.9\% more likely to be matched to a female examiner, converting the logistic regression coefficient for an all-female team to a percentage change from Table \ref{femHomophily}. 
Furthermore, we find that women examiners are more likely to have less examination experience, as depicted in Table \ref{examExperienceRegFemExaminer}. 
The regression results show that inferred examiner gender significantly correlates with examiner experience. 
Furthermore, in a two sample Kolmogorov-Smirnov test, we find that the examination experience of men and women examiners are significantly different ($p<0.001$). 
Gender homophily is one possible explanation for why women inventors are likely to be assigned patent examiners with less experience and thus be confronted with the potential biases and penalties associated with this that manifest in a gendered manner. 

\begin{table}[!htbp]
 \caption{\textbf{Women examiners have less experience overall.} Linear regression results with fixed effects for year, team size, and CPC code.} \label{examExperienceRegFemExaminer}
 \begin{center}

 \resizebox{1\width}{!}{{
\def\sym#1{\ifmmode^{#1}\else\(^{#1}\)\fi}
\begin{tabular}{l*{1}{c}}
\toprule
         DV = Examiner Experience            &\multicolumn{1}{c}{(US Patent Applications)}         \\
\midrule
              &                     \\
Female Examiner           &        -30.222\sym{***}\\
                    &    (0.623)         \\

\addlinespace
Year FE          &     Y \\
\addlinespace
Team Size FE          &     Y \\
\addlinespace
CPC Code FE          &     Y \\
\addlinespace

\midrule
Observations        &      1,338,832        \\
\(R^{2}\)    &         0.3052       \\
\bottomrule
\multicolumn{2}{l}{\footnotesize Standard errors in parentheses}\\
\multicolumn{2}{l}{\footnotesize \sym{*} \(p<0.05\), \sym{**} \(p<0.01\), \sym{***} \(p<0.001\)}\\
\end{tabular}
}
}

\end{center}

\end{table}

\section{Fields where Women More Likely to Invent Given Less Time for Examination}

\begin{figure}[!ht]
\centering
\includegraphics[width=1\textwidth]{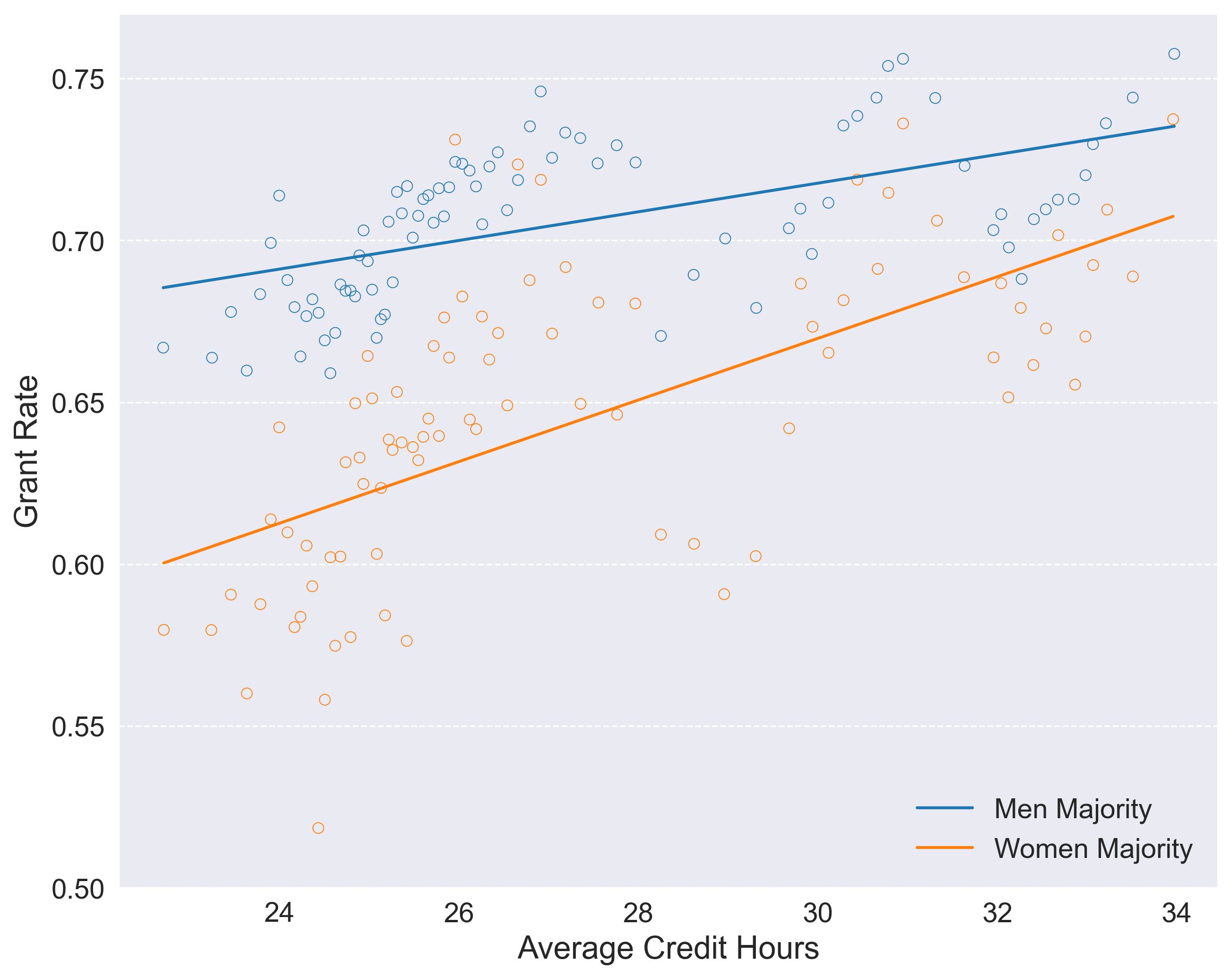}
\caption{\textbf{Average grant rate increases for unconventional patent applications with increased credited examination time provided.} This depicts a binned scatterplot (85 quantiles) of the average credit hours given to examine unconventional patent applications and grant rate by innovator gender with regression lines included that have controls for year, team size, examiner gender, and examiner experience quantile. There is a higher grant rate with more time given for patent examination for both men and women innovators, but women innovators have lower grant rates overall for the same number of credit hours. }

\label{fig:FOIcreditHours}
\end{figure}

Through a Freedom of Information Act data request, we obtained data from the USPTO on how many credit hours each CPC code is assigned for patent examiners.
This represents the amount of credited time a patent examiner gets for an application in a particular CPC code. 
We find that technical areas where women inventor’s propensity to patent is high receive on average significantly lower credit hours than areas men patent in by comparing patent applications by women and men majority teams and the credit hours assigned to these applications in those areas (T = -11.37, $p < 0.0001$).  
We find that for unconventional applications, the average grant rate increases with a higher number of credit hours for both men and women inventors, as depicted in Figure \ref{fig:FOIcreditHours}, which is a binned scatterplot for unconventional applications of average credit hours and grant rate by gender with regression lines for each group depicted. 
However, women innovators have an overall lower grant rates across all levels of credit hours than men.

We also find that the areas where women are more likely to invent are given less time for examination, as depicted in Table \ref{foiRegResults}.
Here, we see that being a women majority inventor team has a negative relationship with the average number of credit hours while controlling for year and team size. 
An innovation being unconventional has a positive coefficient, and the interaction between gender and unconventionality has a negative coefficient, suggesting that when women invent in ways that push the boundaries of convention, their innovations are given even less time for examination. 
This suggests that the fields or technical areas women are innovating in are given less time for patent examination. 
Furthermore, we find that being a women examiner negatively predicts receiving higher credit hours for examination, as depicted in Table \ref{foiaExaminerGender}, and the gender difference is especially pronounced when examining unconventional innovations.
This may exacerbate the tendency of patent examiners to make inaccurate assessments.

\begin{table}[!htbp]
 \caption{\textbf{Areas where women are more likely to invent are given less time for examination, and this exacerbates the tendency to make inaccurate assessments.} Credit hours assigned to CPC codes. There are fixed effects for year and team size.} \label{foiRegResults}
 \begin{center}
 
 \resizebox{1\width}{!}{{
\def\sym#1{\ifmmode^{#1}\else\(^{#1}\)\fi}
\begin{tabular}{l*{1}{c}}
\toprule
         DV = Average Credit Hours            &\multicolumn{1}{c}{(US Patent Applications)}         \\
\midrule
              &                     \\
Women Majority Inventors           &        -5.25e-2\sym{***}\\
                    &    (9.62e-3)         \\

\addlinespace
Unconventionality &   1.84e-3\sym{***}\\
                    &  (1.44e-5)         \\
\addlinespace
Women Majority Inventors         $\times$ Unconventionality &        -4.30e-4\sym{***}\\
                    &    (3.67e-5)         \\

\addlinespace

Year FE          &     Y \\
\addlinespace
Team Size FE          &     Y \\
\addlinespace

\midrule
Observations        &      1,301,227     \\
 \(R^{2}\)    &        0.0302      \\
\bottomrule
\multicolumn{2}{l}{\footnotesize The base case for the indicator of Women Majority Inventors is Men Majority Inventors.}\\
\multicolumn{2}{l}{\footnotesize Standard errors in parentheses}\\
\multicolumn{2}{l}{\footnotesize \sym{*} \(p<0.05\), \sym{**} \(p<0.01\), \sym{***} \(p<0.001\)}\\
\end{tabular}
}
}

\end{center}

\end{table}

\begin{table}[!htbp]
 \caption{\textbf{Women examiners are given less time for examination, especially pronounced for unconventional innovations, and this exacerbates the tendency to make inaccurate assessments.} Credit hours assigned to CPC codes. There are fixed effects for year and team size.} \label{foiaExaminerGender}
 \begin{center}
 
 \resizebox{1\width}{!}{{
\def\sym#1{\ifmmode^{#1}\else\(^{#1}\)\fi}
\begin{tabular}{l*{1}{c}}
\toprule
         DV = Average Credit Hours            &\multicolumn{1}{c}{(US Patent Applications)}         \\
\midrule
              &                     \\
Female Examiner           &       -3.88e-1\sym{***}\\
                    &    (6.19e-3)         \\
Unconventionality          &       1.87e-3\sym{***}\\
                    &    (1.66e-5)         \\
Female Examiner   x  Unconventionality        &       -4.79e-4\sym{***}\\
                    &    (2.68e-5)         \\                    

\addlinespace
Year FE          &     Y \\
\addlinespace
Team Size FE          &     Y \\
\addlinespace

\midrule
Observations        &     1,340,367    \\
\(R^{2}\)    &          0.0331  \\
\bottomrule
\multicolumn{2}{l}{\footnotesize Standard errors in parentheses}\\
\multicolumn{2}{l}{\footnotesize \sym{*} \(p<0.05\), \sym{**} \(p<0.01\), \sym{***} \(p<0.001\)}\\
\end{tabular}
}
}

\end{center}

\end{table}

These findings suggest a possible mechanism for how the gender penalty in patent examination manifests and another policy intervention to combat lower grant rates for women innovators, who might be innovating in fields given fewer credit hours for examination and experiencing lower grant rates, especially when they engage in unconventional innovation. 
A policy intervention would be to assign these fields higher credit hours at the USPTO to possibly combat lower grant rates that manifest in a gendered manner.

\section{Gender Disparities in Examiner Survivorship across Fields}

\begin{figure}[!ht]
\centering
\includegraphics[width=1\linewidth]{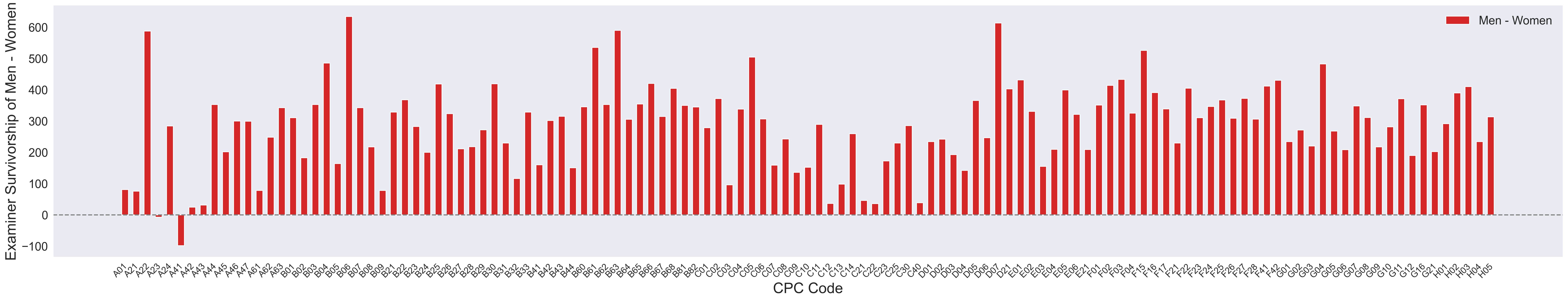}
\caption{\textbf{Examiner survivorship difference of men minus women examiners by CPC code.} The y-axis represents the difference between men and women for the quantity (Particular Gender) Examiner Experience * Proportion of (Particular Gender) Examiners in CPC Code. Values above the gray dashed line at zero represent fields where men have higher survivorship than women, and values below zero represent fields where women have higher survivorship than men. }
\label{fig:survivorship}
\end{figure}

We calculate a metric to understand the field-level variation between CPC codes of gender differences, namely how many examiners of a particular gender are in a field weighted by their average experience in that field.
This is an index of examiner survivorship, where we calculate the quantity of the average examiner experience of men multiplied by the proportion of men examiners in a particular CPC code minus the same quantity but for women examiners. 
When this index is above zero, men have higher survivorship in a particular CPC code than women, and when the index is below zero, women have higher survivorship than men in a particular CPC code. 
Figure \ref{fig:survivorship} shows that in most CPC codes (all but two), men examiners have higher survivorship than women examiners, and only in two fields do women have higher survivorship than men, namely A41 and A23 (very small difference in survivorship for A23).
This demonstrates a pervasive gender gap in examiner survivorship.
These two fields of A41 and A23 are ``wearing apparel'', and ``	foods or foodstuffs" respectively, which some may consider stereotypically ``feminine" fields, further exacerbating that where women do have higher survivorship is only in fields where they are stereotyped to be more feminine. 

\section{UK and Canada Confirmatory Findings and Analysis of Innovation Glass Ceiling}
The USPTO analysis was replicated with the UK and Canada patent application data using similar measures where they exist. Note that the UK and Canada data is less detailed than the USPTO data, which is our richest source of data. 
We find evidence of gender gaps in patent acceptance in the UK and Canada as well.

In a regression model predicting patent acceptance with the US, UK, and Canada patent application data, we find that not only is being a women majority inventor team negatively predictive and unconventionality as a main effect is positive, but the interaction term between unconventionality and women majority inventor team is significant and negative. 
This can be seen in the first ``All Countries" column of Table \ref{regAllCountry}. 
This regression has fixed effects for country, CPC code, team size, and application year.

\begin{table}[!htbp]
  \caption{\textbf{Confirmatory phenomenon of innovation glass ceiling internationally: females are penalized and especially face penalties when engaging in unconventional work.} International Patent Application Logistic Regression Results. There are fixed effects for year, team size, CPC code, and country for the ``All Countries" results.} \label{regAllCountry}
 \begin{center}
 
 \resizebox{1\width}{!}{

{
\def\sym#1{\ifmmode^{#1}\else\(^{#1}\)\fi}
\begin{tabular}{l*{1}{c}}
\toprule
      DV = Patent Acceptance                &\multicolumn{1}{c}{(All Countries)}         \\
\midrule
             &                     \\
Women Majority Inventors           &      -2.36e-1\sym{***}\\
                    &    (7.18e-3)         \\
\addlinespace
Unconventionality             &    2.08e-4\sym{***}\\
                    & (1.23e-5)         \\
\addlinespace
Women Majority Inventors $\times$ Unconventionality&   -1.54e-4\sym{***}\\
                    &  (2.69e-5)         \\
Year FE          &     Y \\
\addlinespace
Team Size FE          &     Y \\
\addlinespace
CPC Code FE          &     Y \\
\addlinespace
Country FE          &     Y \\
\addlinespace
\midrule
Observations        &   1,414,158        \\
Pseudo \(R^{2}\)    &       0.0532   \\
\bottomrule
\multicolumn{2}{l}{\footnotesize The base case for the indicator of Women Majority Inventors is Men Majority Inventors.}\\
\multicolumn{2}{l}{\footnotesize Standard errors in parentheses}\\
\multicolumn{2}{l}{\footnotesize \sym{*} \(p<0.05\), \sym{**} \(p<0.01\), \sym{***} \(p<0.001\)}\\
\end{tabular}
}
}

\end{center}

\end{table}

For the UK and Canada, because we have access to fewer applications, we rely on the cumulative CPC code z-score network from the US patent application data to compute the unconventionality of UK and Canadian applications. 
This is done using Equations \eqref{eq:atyp} and \eqref{eq:atypInv} given the CPC codes in the applications from the UK and Canada while using the cumulative z-score network in the corresponding time subset of the US application data.

We find further evidence of differences in acceptance rates between men and women when evaluating unconventional applications in all three countries' data. To isolate applications that are likeliest to present examiners with both conventional and unconventional combinations of technical fields, we focus on applications with more than one CPC subclass assignments. This will include some applications with typical combinations of CPC codes that fit into examiners' existing understandings of the technologies they specialize in, as well as other less typical combinations that are harder for examiners to evaluate and take longer to assess \cite{whalen_boundary_2018, harhoff2009duration}.

\section{High Potential Innovation Predicted by Unconventionality}

Research shows that patent citations are useful measures of an invention’s value \cite{trajtenberg1990penny}. They correspond to the patenting firm’s market value \cite{hall2005market}, expert assessment of the value of the underlying invention \cite{albert1991direct, harhoff1999citation}, and objective measures of the effectiveness of the patented technology \cite{moser2015patent}.

There are several ways to assess and estimate the future impact of a patent that has been granted. 
One of these measures is through citation count, which we predict the citation count in regression models, where this citation count represents the number of citations a granted patent has garnered in 8 years. 
We find that unconventionality is positively predictive of citation count, while the proportion of female inventors is negatively predictive. This is depicted in Table \ref{citationImpact}. We also find that unconventionality is positively predictive of being a top 5\% cited patent in Table \ref{citationImpactLogit}. 

\begin{table}[!htbp]
 \caption{\textbf{Unconventionality Predictive of Future Citation Impact.} Citation count regression results with fixed effects for grant year, team size, and CPC code. } \label{citationImpact}
 \begin{center}

     \resizebox{1\width}{!}{

{
\def\sym#1{\ifmmode^{#1}\else\(^{#1}\)\fi}
\begin{tabular}{l*{1}{c}}
\toprule
      DV = Citation Count               &\multicolumn{1}{c}{(US Granted Patents)}         \\
\midrule
             &                     \\
Women Majority Inventors           &      -1.26\sym{***}\\
                    &    (1.35e-1)         \\
\addlinespace
Unconventionality             &    6.40e-3\sym{***}\\
                    & (3.24e-4)         \\
\addlinespace
Examiner Experience             &    -5.74e-4\sym{**}\\
                    & (1.90e-4)         \\
                    
\addlinespace
Female Examiner             &    -3.56e-1\sym{***}\\
                    & (8.53e-2)         \\
                    
\addlinespace
Average Inventor Experience             &    7.25e-4\\
                    & (4.92e-4)         \\
\addlinespace
Big Entity             &   2.54e-2\\
                    & (9.75e-2)         \\
                    
\addlinespace
Grant Year FE          &     Y \\
\addlinespace
Team Size FE          &     Y \\
\addlinespace
CPC Code FE          &     Y \\
\addlinespace
\midrule
Observations        &    182,890      \\
\(R^{2}\)    &         0.0187 \\
\bottomrule
\multicolumn{2}{l}{\footnotesize This is for granted patents accepted in 2011 and previously.}\\
\multicolumn{2}{l}{\footnotesize The base case for the indicator of Women Majority Inventors is Men Majority Inventors.}\\
\multicolumn{2}{l}{\footnotesize Standard errors in parentheses}\\
\multicolumn{2}{l}{\footnotesize \sym{*} \(p<0.05\), \sym{**} \(p<0.01\), \sym{***} \(p<0.001\)}\\
\end{tabular}
}
}

\end{center}

\end{table}

\begin{table}[!htbp]
 \caption{\textbf{Unconventionality Predictive of Future Top 5\% Cited Patents.} Logistic regression results for being a top cited patent with fixed effects for grant year, team size, and CPC code.} \label{citationImpactLogit}
 \begin{center}

     \resizebox{1\width}{!}{

{
\def\sym#1{\ifmmode^{#1}\else\(^{#1}\)\fi}
\begin{tabular}{l*{1}{c}}
\toprule
      DV = Top 5\% Cited Patents (Indicator)              &\multicolumn{1}{c}{(US Granted Patents)}         \\
\midrule
             &                     \\
Women Majority Inventors           &      -4.10e-1\sym{***}\\
                    &    (4.75e-2)         \\
\addlinespace
Unconventionality             &    1.71e-3\sym{***}\\
                    & (1.14e-4)         \\
\addlinespace
Examiner Experience             &    -2.66e-4\sym{***}\\
                    & (5.99e-5)         \\
                    
\addlinespace
Female Examiner             &    -8.83e-2\sym{***}\\
                    & (2.64e-2)         \\
                    
\addlinespace
Average Inventor Experience             &   2.66e-4\sym{*}\\
                    & (1.18e-4)         \\
\addlinespace
Big Entity             &    -4.07e-2\\
                    & (2.96e-2)         \\
                    
\addlinespace
Grant Year FE          &     Y \\
\addlinespace
Team Size FE          &     Y \\
\addlinespace
CPC Code FE          &     Y \\
\addlinespace
\midrule
Observations        &    182,406      \\
Pseudo \(R^{2}\)    &        0.0400\\
\bottomrule
\multicolumn{2}{l}{\footnotesize This is for granted patents accepted in 2011 and previously. A top 5\% cited patent has 27 or more citations.}\\
\multicolumn{2}{l}{\footnotesize The base case for the indicator of Women Majority Inventors is Men Majority Inventors.}\\
\multicolumn{2}{l}{\footnotesize Standard errors in parentheses}\\
\multicolumn{2}{l}{\footnotesize \sym{*} \(p<0.05\), \sym{**} \(p<0.01\), \sym{***} \(p<0.001\)}\\
\end{tabular}
}
}

\end{center}

\end{table}

Another way to quantify impact is by tracking patent owners' continued investment in their patents. Granted patents need to pay fees every four years to maintain their patent, deemed a `maintenance fee'. 
We track these event codes for maintenance fees in the data. 
Presumably, valuable inventions will be paying maintenance fees to maintain their patents. 
We find that in predicting an indicator variable of whether any maintenance fee payment was made at any time (for granted applications), the proportion of female inventors is a negative coefficient while unconventionality is positively predictive, as depicted in Table \ref{maintenanceAnyFee}. 

\begin{table}[!htbp]
 \caption{\textbf{Unconventionality Predictive of Patent Maintenance.} Logistic regression results for a patent being paid maintenance fees with fixed effects for grant year, team size, and CPC code.} \label{maintenanceAnyFee}
 \begin{center}

    \resizebox{1\width}{!}{

{
\def\sym#1{\ifmmode^{#1}\else\(^{#1}\)\fi}
\begin{tabular}{l*{1}{c}}
\toprule
      DV = Any Maintenance Fee Paid (Indicator)               &\multicolumn{1}{c}{(US Granted Patents)}         \\
\midrule
             &                     \\
Women Majority Inventors           &      -1.04e-1\sym{***}\\
                    &    (1.49e-2)         \\
\addlinespace
Unconventionality             &    1.49e-4\sym{***}\\
                    & (2.62e-5)         \\
\addlinespace
Examiner Experience             &    5.35e-5\sym{***}\\
                    & (1.61e-5)         \\
                    
\addlinespace
Female Examiner             &    -3.99e-3\\
                    & (9.91e-3)         \\
                    
\addlinespace
Average Inventor Experience             &    -7.89e-4\sym{***}\\
                    & (5.83e-5)         \\
\addlinespace
Big Entity             &    2.42e-1\sym{***}\\
                    & (1.06e-2)         \\
                    
\addlinespace
Grant Year FE          &     Y \\
\addlinespace
Team Size FE          &     Y \\
\addlinespace
CPC Code FE          &     Y \\
\addlinespace
\midrule
Observations        &    434,348     \\
Pseudo \(R^{2}\)    &      0.0110 \\
\bottomrule
\multicolumn{2}{l}{\footnotesize This is for patents with grant year at or below 2015 to allow time for accrual of maintenance fees.}\\
\multicolumn{2}{l}{\footnotesize The base case for the indicator of Women Majority Inventors is Men Majority Inventors.}\\
\multicolumn{2}{l}{\footnotesize Standard errors in parentheses}\\
\multicolumn{2}{l}{\footnotesize \sym{*} \(p<0.05\), \sym{**} \(p<0.01\), \sym{***} \(p<0.001\)}\\
\end{tabular}
}
}

\end{center}

\end{table}

Both our measures of impact suggest unconventional inventions correspond and predict future value through both citations garnered and patent maintenance.

\section{Patent Value Estimates and Quantified Lost Value from Gender Gap in Success Rate}

Patent value is difficult to estimate precisely, and highly skewed. We estimate the monetary value of a patent by averaging values from previous estimates.
Previous literature estimates of value averaged across different technical fields is \$101,850 (1992 USD) per patent \cite{fischer2014testing}.
Previous estimates of value averaged across US entity types is \$107,557 (1992 USD) \cite{bessen2008value}. We use the average across these as a rough estimate of per patent value: \$104,703.5 (1992 USD). 

We estimate the number and value of unconventional “lost patents” that would have been pioneered by women by inferring how many more unconventional patents would be granted if women and men had the same grant rate (for majority teams). This is done by multiplying the number of women’s unconventional applications (33,761) by the difference in grant rate between men and women in this type of innovation (men success rate 70.55\%, women success rate 63.92\%, the gender difference is approximately 6.63\%), yielding an additional approximately 2,238 lost patents by women innovators that may have been patented and circulated in the market. If we assume average patent value from prior estimates in the literature \cite{bessen2008value, fischer2014testing}, this equates to women inventors not being granted intellectual property rights valued at over \$234 million (estimated by multiplying the number of lost patents by their average value of \$104,703.5 (1992 USD)).

\subsection{Estimating Effect of Higher Experienced Examiner Assignment when Evaluating Women's Unconventional Inventions}

We estimate the increase in grant rate for women's unconventional patents if they were assigned more high experienced examiners rather than low experience examiners. To do this, we take the grant rate for high experience women examiners evaluating women majority inventor teams doing unconventional work (70.4\% grant rate). We then subtract the grant rate for low experience women examiners evaluating women majority inventor teams doing unconventional work (44.3\% grant rate). This difference is a 26.1\% difference in grant rate if women inventors in this category were assigned to women examiners with higher experience. If we take the reassignment of women inventors doing unconventional work from low experience to high experience women examiners at just 50\%, then we multiply the grant rate difference with the reassignment rate to get that there would be 13\% increase in granted unconventional work from women inventors with a moderate change in examiner experience assignment to women.

\end{document}